\documentclass[utf8]{FrontiersinHarvard} 

\usepackage{url,hyperref,lineno,microtype,subcaption,graphicx}
\usepackage[onehalfspacing]{setspace}

\def\keyFont{\fontsize{8}{11}\helveticabold }
\def\firstAuthorLast{Gonz\'alez-P\'erez {et~al.}} 
\def\Authors{Jos\'e Nicol\'as Gonz\'alez-P\'erez\,$^{1,*}$, Marco Mittag\,$^{1}$, J\"urgen H. M. M. Schmitt\,$^{1}$, Klaus-Peter Schr\"oder\,$^{2}$, Dennis Jack $^{2}$, Gregor Rauw\,$^{3}$, and Ya\"el Naz\'e\,$^{3}$}
\def\Address{$^{1}$Hamburger Sternwarte, Universit\"at Hamburg, Hamburg, Germany \\
$^{2}$Departamento de Astronom\'{\i}a, Universidad de Guanajuato, Guanajuato, Mexico \\
$^{3}$Groupe d'Astrophysique des Hautes Energies, STAR, Universit\'e de Li\`ege, Li\`ege, Belgium}
\def\corrAuthor{Gonz\'alez-P\'erez, J. N.}

\def\corrEmail{st1h317@hs.uni-hamburg.de}

\begin{document}
\onecolumn
\firstpage{1}

\title[Eight years of TIGRE]{Eight years of TIGRE robotic spectroscopy: Operational experience and selected scientific results
} 

\author[\firstAuthorLast ]{\Authors} 
\address{} 
\correspondance{} 

\extraAuth{}

\maketitle

\begin{abstract}

\section{}

TIGRE (Telescopio Internacional de Guanajuato Rob\'otico Espectrosc\'opico) has been operating in fully robotic mode in the Observatory La Luz (Guanajuato, Mexico) since the end of 2013. With its sole instrument, HEROS, an \'echelle spectrograph with a spectral resolution R$\sim$20000, TIGRE has collected more than 48000 spectra of 1151 different sources with a total exposure time of more than 11000 hours in these eight years. Here we briefly describe the system and the upgrades performed during the last years. We present the statistics of the weather conditions at the La Luz Observatory, emphasizing the characteristics that affect the astronomical observations. We evaluate the performance and efficiency of TIGRE, both optical and operational, and describe the improvements of the system implemented to optimize the telescope's performance and meet the requirements of the astronomer in terms of timing constraints for the observations and the quality of the spectra. We describe the actions taken to slow down the optical efficiency loss due to the aging of the optical surfaces as well as the upgrades of the scheduler and the observing procedures to minimize the time lost due to interrupted observations or observations that do not reach the required quality. Finally, we highlight a few of the main scientific results obtained with TIGRE data.

\tiny
 \keyFont{ \section{Keywords:} Automated telescopes, Instrumentation: spectrographs, Techniques: spectroscopic, Atmospheric effects, Stars: activity, Stars: massive, Novae, Supernovae} 
\end{abstract}

\section{Introduction}

Robotic telescopes have gained considerable importance over the last two decades \citep{2010AdAst2010E..60C}, and so the advent of robotic astronomy has transformed observational astronomy due to the increased economic and operational efficiency. In addition, robotic telescopes allow different modes of observation that are not possible or are only very difficult to achieve with ``normal'' telescopes. These include rapid alerts, long-term monitoring, phase-constrained observations, or simultaneous observations with other observatories. 

In the list of robotic telescopes maintained by F. V. Hessmann\footnote{\url{http://www.astro.physik.uni-goettingen.de/~hessman/MONET/links.html}; last actualization from 2016}, one finds more than eighty robotic telescopes in operation and more than forty in commissioning, under construction, or planning. While most of these telescopes are small, a few tens have a diameter larger than 1\,m. Even a 4\,m diameter telescope, the NRT, will be starting operating in La Palma in the next few years \citep{2019AN....340...40G}. The scientific goals of the robotic telescopes cover the detection of near-Earth asteroids (such as LINEAR, \citealp{2000Icar..148...21S}), the study of exoplanets (e.g., SuperWASP, \citealp{2006Ap&SS.304..253P}), to the detection of rapid transients such as $\gamma$-ray Bursts (e.g., BOOTES, \citealp{2012ASInC...7..313C}). Only a few of these telescopes are exclusively dedicated to high-resolution spectroscopy, such as STELLA-1 \citep{2004AN....325..527S}, SONG \citep{2019PASP..131d5003F}, or TIGRE \citep{2014AN....335..787S}, the topic of this paper.

\begin{figure} 
\begin{center}
\includegraphics[angle=-90,width=8cm]{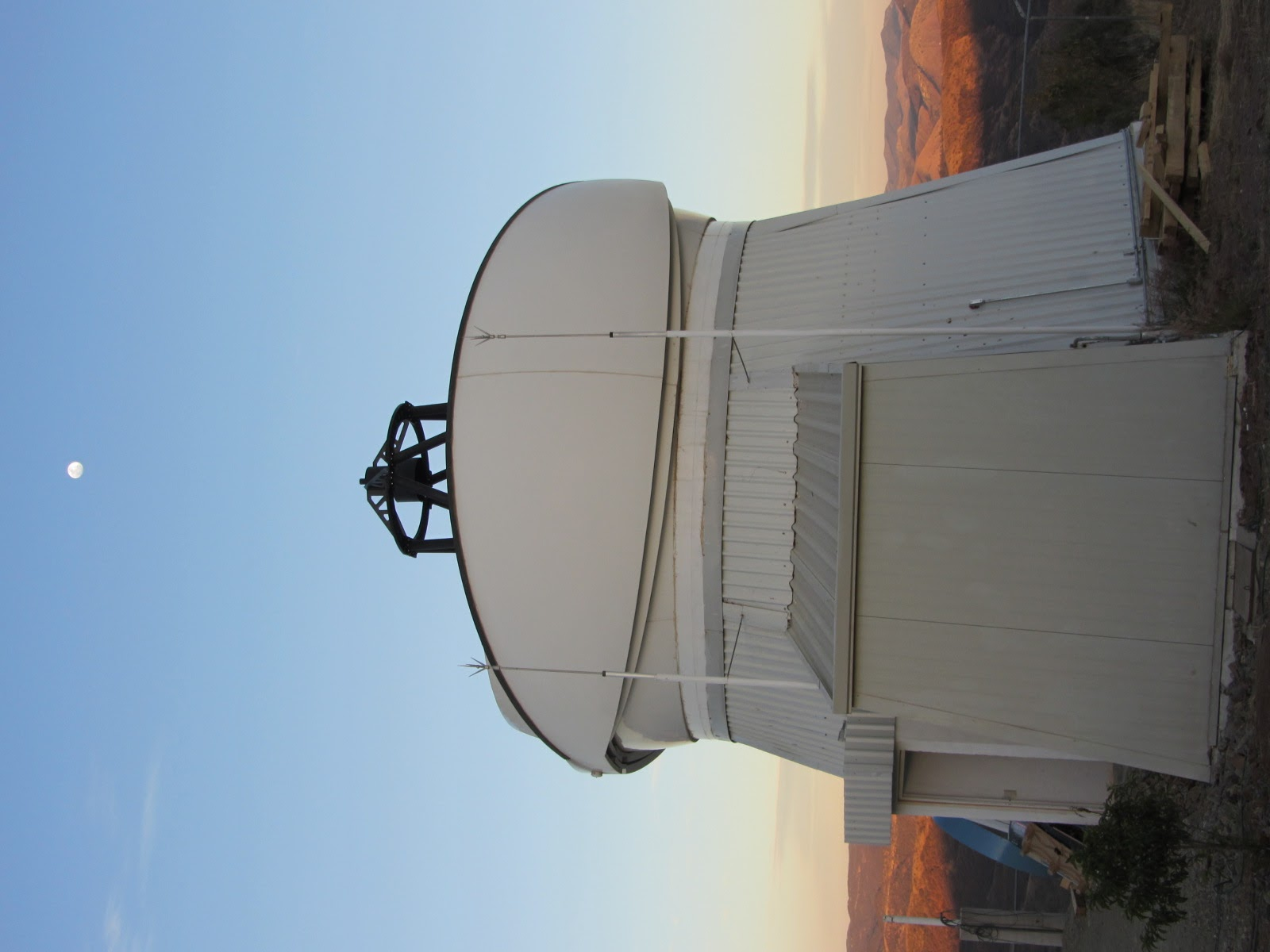}
\end{center}
\caption{The TIGRE facility at La Luz Observatory.}\label{fig:foto}
\end{figure}

The TIGRE (Telescopio Internacional de Guanajuato Rob\'otico Espectrosc\'opico) project is a collaboration between Hamburg Observatory (Germany), and the universities of
Guanajuato (Mexico) and Li\`ege (Belgium). The telescope was installed in 2013 in its final location, the La Luz Observatory in central Mexico (see Fig.~\ref{fig:foto}, which gives an impression of the
telescope building and its surroundings), but it was first delivered to Hamburg, where it was used to test and develop the final system \citep{2010PhDT.......503M}. Initially, TIGRE was 
designed to study the stellar activity of cool stars using the CaII H\&K line cores, but over the years it has widened its scientific goals to other fields of stellar astronomy, like hot stars, novae, or binaries. 

In this paper, we describe our experience with the first eight years of fully-automatic operation of TIGRE. We summarize the characteristics of the telescope, instrument, and software in Sect.~\ref{Sect:system}. We also describe in Sect.~\ref{Sect:observatory} the La Luz Observatory and report the weather, seeing, and photometric statistics of the observatory. In Sect.~\ref{Sect:perfor}, we discuss the different aspects of the efficiency of TIGRE, from the optical to the operational point of view. Finally, in Sect.~\ref{Sect:Sci}, we outline some of the main scientific results obtained with TIGRE.

\section{System overview}
\label{Sect:system}

The TIGRE telescope was manufactured by Halfmann Teleskoptechnik GmbH. The Cassegrain-Nasmyth optics of the telescope was built by Carl Zeiss Jena and made of Zerodur. The primary mirror (M1) has a diameter of 1.2\,m with a focal ratio of f/3. With the secondary mirror, the focal distance of the system is 9.6\,m with a focal ratio of f/8. The telescope has two Cassegrain-Nasmyth foci that can be accessed by changing the position of the flat M3 mirror, although currently only one of the foci is in use. Since M1 is relatively thin, it has actuating cell support with a total of 30 static levers, 18 at the back of the mirror arranged in two rings, and 12 at the edge of the mirror.

The compact Alt/Az mount has an hydraulic bearing in the azimuth axis and a mechanical bearing in the elevation axis. The high precision encoders allow a high slewing velocity of 5$^{\circ}$s$^{-1}$ and high precision pointing and tracking. Using an extended pointing model, with 17 parameters instead of the standard 11 for a Nasmyth telescope, we could improve the pointing accuracy to $<5''$ \citep{2008PASP..120..425M}.

The only focal-plane instrument of TIGRE is the Heidelberg Extended Range Optical Spectrograph (HEROS), which is located in a thermally controlled room. Its spectral resolution is $\lambda/\Delta\lambda \sim 20000$, and a beam-splitter divides the spectrum into two channels with wavelength ranges of 3740-5740\,\AA\, and 5770-8830\,\AA. The CCD cameras of each channel are Andor iKonL CCDs with E2V 2K$\times$2K chips and a pixel size of 13.5\,$\mu$m. These CCDs have very low dark noise and high quantum efficiency and are optimized for the wavelength range of their respective channels and reach quantum efficiencies of 80\% at 3900\,\AA, 93\% at 6000\,\AA, and 50\% at 8500\,\AA. The spectrograph is fed by a 15\,m fused silica multimode fiber with a core diameter of 50\,$\mu$m. Microlenses are attached to both ends of the fiber to adapt the focal ratios of the spectrograph and telescope.

The adapter, located in one of the Cassegrain-Nasmyth foci, contains the acquisition and guiding unit, the calibration unit, and feeds the light of the stars or the calibration lamps into the optical fiber. The diameter of the microlens at the fiber entrance is 150\,$\mu$m which translates to $\sim3''$ on the sky. A pellicle beam-splitter in the adapter deviates 8\% of the starlight to the guiding camera, and the remaining 92\% enter the fiber. Although losing a small fraction of the light, this procedure allows us to accurately monitor the seeing and the transparency changes of the atmosphere in real-time, which is very useful to estimate the necessary exposure times (see below). 

The guiding camera is an Atik 420m CCD with a field of view of $2.6'\times1.9'$. A set of neutral density filters helps to fix the exposure time of the guiding to 10\,s regardless of the star's brightness. Also, the adapter incorporates a tungsten lamp and a hollow cathode ThAr lamp for flat-field and wavelength calibrations, respectively.

The design of the building is very compact (cf. Fig.~\ref{fig:foto}); the clamshell dome, with a diameter of only 6\,m, thanks to the small size of the mount, was manufactured by Astrohaven Enterprises. The telescope is mounted over a pier 3.5\,m above ground to avoid most of the dust raised by wind during the dry season. The building has two extensions, one of which is the air-conditioned spectrograph room, and the second contains the electronic cabinet and the hydraulic unit of the telescope.

The monitoring of the weather parameters is absolutely essential for the system's health. We monitor the weather conditions at La Luz Observatory using three different weather stations and an independent rain sensor. Two of the weather stations are placed on a $\sim$6\,m high pole, together with a GPS antenna that provides the accurate time necessary for the telescope pointing and tracking. Additionally, we use an all-sky camera to assess the cloudiness above the observatory independently. The most critical parameters are humidity, rain, wind, and sky temperature. Each of these four parameters are provided by more than one weather station, so the system can continue monitoring these parameter in case of failure of one of the devices.

The sky temperature device contains an IR photometer pointed to the sky and an air thermometer. It measures the difference between the ambient temperature and the sky temperature. In a cloudy sky, the sky temperature is the temperature at the bottom of the clouds. This temperature difference gives us a rough estimate of the level of cloudiness, so a threshold in this temperature difference can be used to decide when to close due to a cloudy sky. Although this relationship is far from being perfect --it depends on other factors, e.g., humidity-- it is very effective to evaluate the rain risk, thus protecting the telescope's health. 

The software of TIGRE, written in Java under Linux, was designed simple and modular, to make it very easy to substitute or add a device, such as a weather station. Each subsystem of TIGRE is operated by its program that communicates with the main program (CCS) through TCP/IP ASCII messages. Two essential parts of the CCS are the robotic operator and the error handler. The robotic operation of the telescope is fully automatic and is divided into procedures: initialization, calibrations, observation, open, close, and shutdown. Each procedure starts when several conditions are met. These conditions depend on the weather, time of the day, Sun elevation, and whether other procedures have already been performed. Also, we implemented a comprehensive error detection and handling. We are continuously improving the error handler once we gain experience with any upcoming issues. In the event of severe problems, where the system is at risk, the error handler closes and shuts the telescope down and sends an e-mail to the technical staff reporting the issue. 

The scheduler, one of the subsystems, selects the star to be observed from a pool of stars considering a series of multiplicative weights. These weights account for the targets' scientific priority, the star's position in the sky (the scheduler favors stars close to the meridian), the timing requirements of the astronomer, and whether the available time in the night is enough to finish the observation. Also, the scheduler considers Targets of Opportunities (ToOs) that have the highest priority. TIGRE will observe the active ToOs when the weather conditions are favorable and all other requirements are fulfilled.

At the end of the night, once all calibrations are finished, the data are copied to Hamburg, where the automatic reduction pipeline immediately starts. The pipeline is written in IDL and adapted from the REDUCE package \citep{2002A&A...385.1095P} to the TIGRE-HEROS data. It corrects the raw images from bias, automatically calibrates the wavelength, calculates the order positions, and makes an optimal extraction of the spectra. With the flat-field frames taken each night, the pipeline performs the flat field and the blaze function corrections (\citealp{2010AdAst2010E...6M}). Also, it automatically calculates the radial velocity (RV) of the stars \citep{2018AN....339...53M} and several stellar activity indices from the Ca\,{\sc ii} H\&K and infrared triple lines \citep{2016A&A...586A..14H,2016A&A...591A..89M,2017A&A...607A..87M}. Besides, the pipeline produces a series of plots to monitor the CCDs' performance and the reduction quality. TIGRE observes each night one photometric standard star and one RV standard star. These observations are used to remove the residuals from the blaze correction and monitor the instrument's RV stability, respectively. 

TIGar, the TIGRE archive, is the interface between the astronomer and the telescope. It comprises a MySQL database and has access to the FITS files (reduced and raw spectra) obtained by TIGRE. The human interface is a webpage written in PHP. Using this webpage, the astronomer can create or edit a proposal and add the stars to be observed with their requirements. Also, when the observations are finished and reduced, the astronomer can retrieve the data through the same webpage. 

\section{La Luz Observatory}
\label{Sect:observatory}

La Luz Observatory is located about 20\,km from Guanajuato, about 300\,km NW of Mexico City in the high plateau of central Mexico at coordinates 101.32478W and 21.053139N and at an elevation of 2435\,m above sea level. 


Since the observatory is relatively isolated, an internet microwave beam antenna connects to a server on the Cerro de Cubilete around 10 km away. The electricity supply, unfortunately, shows frequent power failures, overvoltages, and surges, often related to thunderstorms and strong winds. A UPS and a power plant operating in the observatory guarantee a regulated power supply.

The climate in the observatory has two very different seasons: a very dry winter and a humid-rainy summer. The conditions in winter are excellent for the observations (see below), but the dust production is relatively high due to the dryness and the type of soil. After the road to the observatory was paved in 2014, the dust contamination has decreased significantly. In the summer, on the contrary, thunderstorms cause many power failures, and lightning strikes can cause severe damage to the system. Although we have a robust lightning protection, with surge protection in the power line and lightning rods, we already had to replace the internet and GPS antennas twice.

In the following subsections, we provide a detailed analysis of the weather conditions and the seeing in the observatory. 

\subsection{Weather conditions}

We have collected the weather data of 2015-2021 from our weather stations and present their statistics in this section. January is the coldest month with an average temperature of 12.5$^\circ$C (the average nightly temperature is 11$^\circ$C). Afterwards, the monthly average temperatures rise continuously until May, when the average temperature reaches 19$^\circ$C (17$^\circ$C at night). Then, the temperatures drop ($<$17$^\circ$C), even in the summer, because the cloudy sky prevents the ground to be excessively heated.

The temperature variations during the night are relatively small, typically below five degrees between June and December and less than seven degrees during the rest of the year. The maximum daily temperature contrast is usually below 10 degrees, and between July and January below 8.5 degrees. A small temperature drop in the night helps to match the telescope and air temperatures and thus reduce the influence of the local air layer in the seeing.

The daily-averaged relative humidity shows seasonal variations, with a minimum in April (around 30\%) and a maximum in September (70\%). From June to October, the humidity is generally high ($>$60\%). The rainiest month is June, followed by July, August, and September in this order.

There are no substantial variations of the monthly-average wind velocities, with values between 2\,m\,s$^{-1}$ and 4\,m\,s$^{-1}$. However, the averages of the maximal-minimal wind daily differences have a maximum of 9.5\,m\,s$^{-1}$ in June. It is higher than 8\,m\,s$^{-1}$ between March and October, indicating that the probability of having adverse winds preventing any observations is much higher during these months. Furthermore, the windy nights are also the nights with larger seeing (see below), which has led us to analyze the frequency of the strong winds. The fraction of the time with wind velocities $>$7.5\,m\,s$^{-1}$ grows from ~0.02 in December and January to 0.11 in June. After June, this fraction remains higher than 5\% until October. Furthermore, windy conditions in summer are ten times more probable when the sky is clear than cloudy.

To finalize this section, we study the periods of time with good weather conditions and how this changes through the year. Good weather conditions are defined by thresholds in some weather parameters: wind, humidity, sky temperature, and rain. We principally set these limits to avoid damages in the system and evaluate when the observations are possible with a high probability. Observations are allowed only for wind velocities lower than 10\,m\,s$^{-1}$. With stronger winds, the telescope starts oscillating, and the dome may be damaged. The maximum permitted humidity is 90\% to avoid condensation on the mirror surfaces. Any rain detection also triggers the procedure to abort the observations and close the dome. 

The last threshold is for the difference between the sky and ambient temperature: we consider the sky as cloudy if this difference is higher than -28$^{\circ}$C. Note that this limit cannot be extrapolated to other observatories because it depends on the location, particularly on the elevation above sea level and the device itself. For the setting of this limit, we have carefully compared the sky temperature values with the all-sky camera images. However, it should be emphasized that the correlation between the sky temperature and the cloudiness is not very tight. The sky temperature is not very sensitive to high clouds; thus, the sky may be covered by high clouds, and the sky temperature is below the threshold by two or three degrees. The robotic operator handles this possibility by estimating the necessary exposure time using the guiding images (see below), or aborting the observation if the star is not detected. 

If any of these limits are exceeded, the observations are aborted or not allowed, and the dome remains closed. After the weather parameters return below the limits, we wait for additional 15 minutes before the observations start again. In this fashion, we ensure stable conditions and avoid opening and closing the dome too often. We link this waiting time to the last bad-weather event in the following analysis.

\begin{figure} 
\begin{center}
\includegraphics[width=12cm]{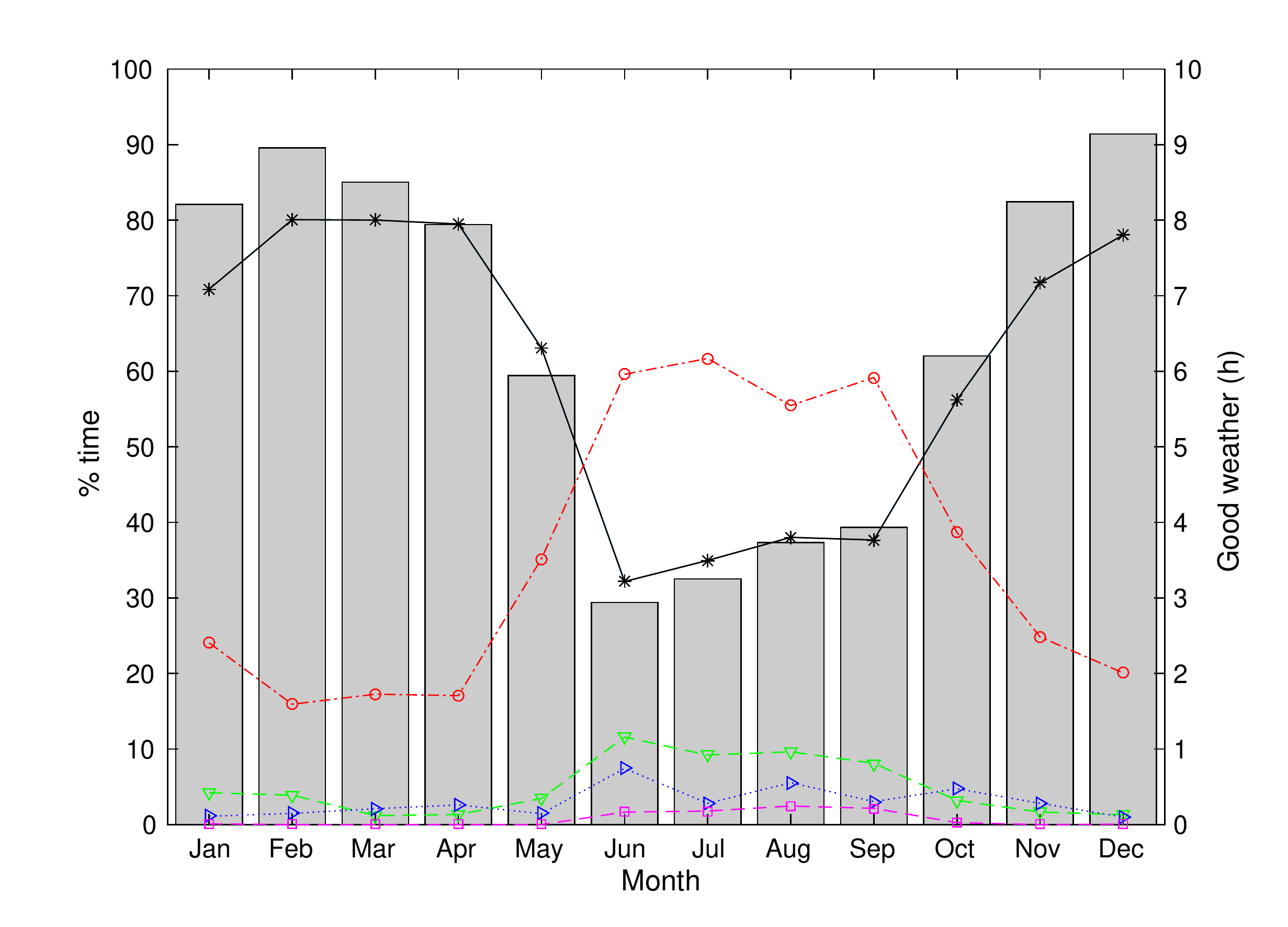}
\end{center}
\caption{Grey bars (right axis) show the average daily number of hours of good weather in each month. The color symbols (left axis) display the percentage of time of good weather (black asterisks) and observations not allowed by clouds (red circles), rain (green downward-pointing triangles), wind (blue right-pointing triangles), and humidity (magenta squares).}\label{fig:badWeather}
\end{figure}

Between 2015 and 2021, 61\% of the available time had good weather, 35\% of the time was cloudy, and it was rainy 5\%, humid 1\%, and windy 3\% of the time. Figure~\ref{fig:badWeather} shows the monthly percentage of the time with good weather. Also, it displays the frequency of the different conditions that prevent the observations. We also provide the average total number of hours per night available for astronomical observations; as obvious from Fig.~\ref{fig:badWeather}, we have typically 8 hours in the winter (November-April) and 3-4 hours in the Summer (June-September)

In contrast to other northern observatories with a worse winter and better summer conditions, we have excellent winter conditions at La Luz Observatory, with more than 70\% of the time with observable conditions between November and April, and reaching 80\% of the time in several months. The wet season starts very fast, with June being the worst month for the observations, with a short transition period in May. Between June and September, only 30-40\% of the time can be used for the observations.

The main reasons for the bad weather are clouds, around 20\% in the winter and 60\% in the wet seasons. On the other hand, rain is present around 10\% of the time in summer. The wind is usually less critical, but as described above, it can be an issue in summer when it is clear.

\subsection{Seeing statistics}

Even for a spectroscopic telescope like TIGRE, good seeing is essential. The diameter of the microlens in front of the fiber entrance is 3$''$, implying that a large amount of photons is lost if the seeing  becomes too large. To compute how many photos are lost by such an aperture, we assume a seeing profile modeled as a Moffat function as follows:

\begin{equation}
F(r) \propto \left[1 + \left(\frac{r}{\alpha}\right)^2 \right]^{-\beta} ,
\end{equation}

\noindent with $\alpha$ and $\beta$ are seeing dependent parameters (FWHM$=2\alpha\sqrt{2^{1/\beta}-1}$). Assuming $\beta=3$, only 25\% of the photons fall inside a centered circle of diameter 3$''$ with a seeing of FWHM=4$''$. This fraction of photons increases to 38\%, 60\%, and 91\% respectively, when the seeing is 3$''$, 2$''$, and 1$''$. Although poor seeing reduces the number of detected photons significantly, it is not necessary to have excellent seeing to collect most of the photons.

\begin{figure} 
\begin{center}
\includegraphics[width=12cm]{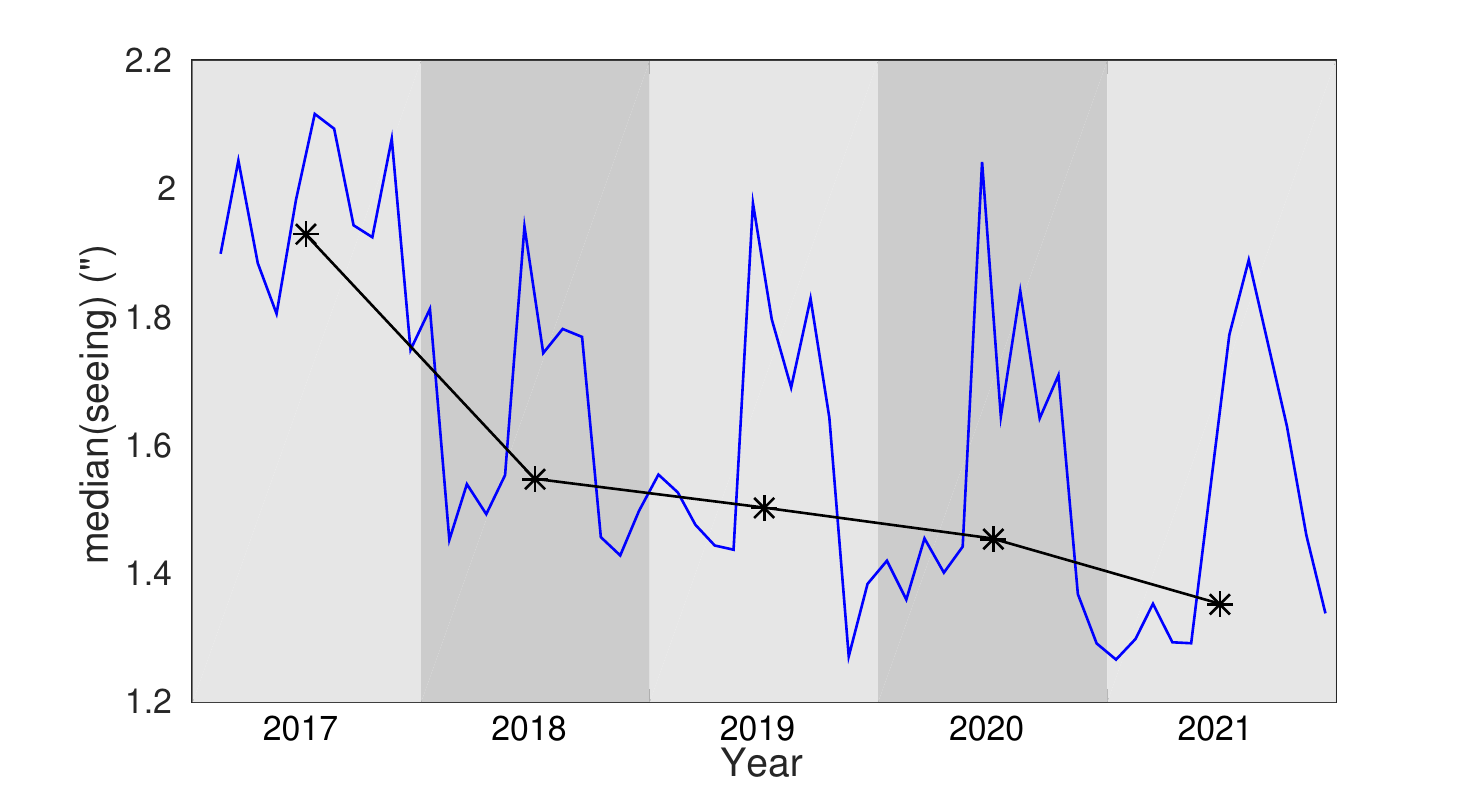}
\end{center}
\caption{Monthly median of the seeing observed by TIGRE between 2017 and 2021 (blue line). The black asterisks and lines show the yearly medians of the seeing.}\label{fig:seeing1_evol}
\end{figure}

We have collected the seeing values from 2017 to 2021 measured by the guiding camera during the acquisition and guiding processes. The scheduler favors the observations close to the meridian, but many observations are nonetheless taken at lower elevations. Seeing does depend on elevation; following the Kolmogorov turbulence model, one expects the seeing to be proportional to the airmass ($X$) as seeing$\propto X^{(2/3)}$ (see e.g., \citealp{2000asop.conf.....S}). Using this correction, we show in Fig.~\ref{fig:seeing1_evol} the evolution of the monthly median seeing at zenith between 2017 and 2021. There is an apparent variation inside a year and also an unexplained and intriguing decline of the yearly-averaged seeing in the last years.

\begin{figure} 
\begin{center}
\includegraphics[width=12cm]{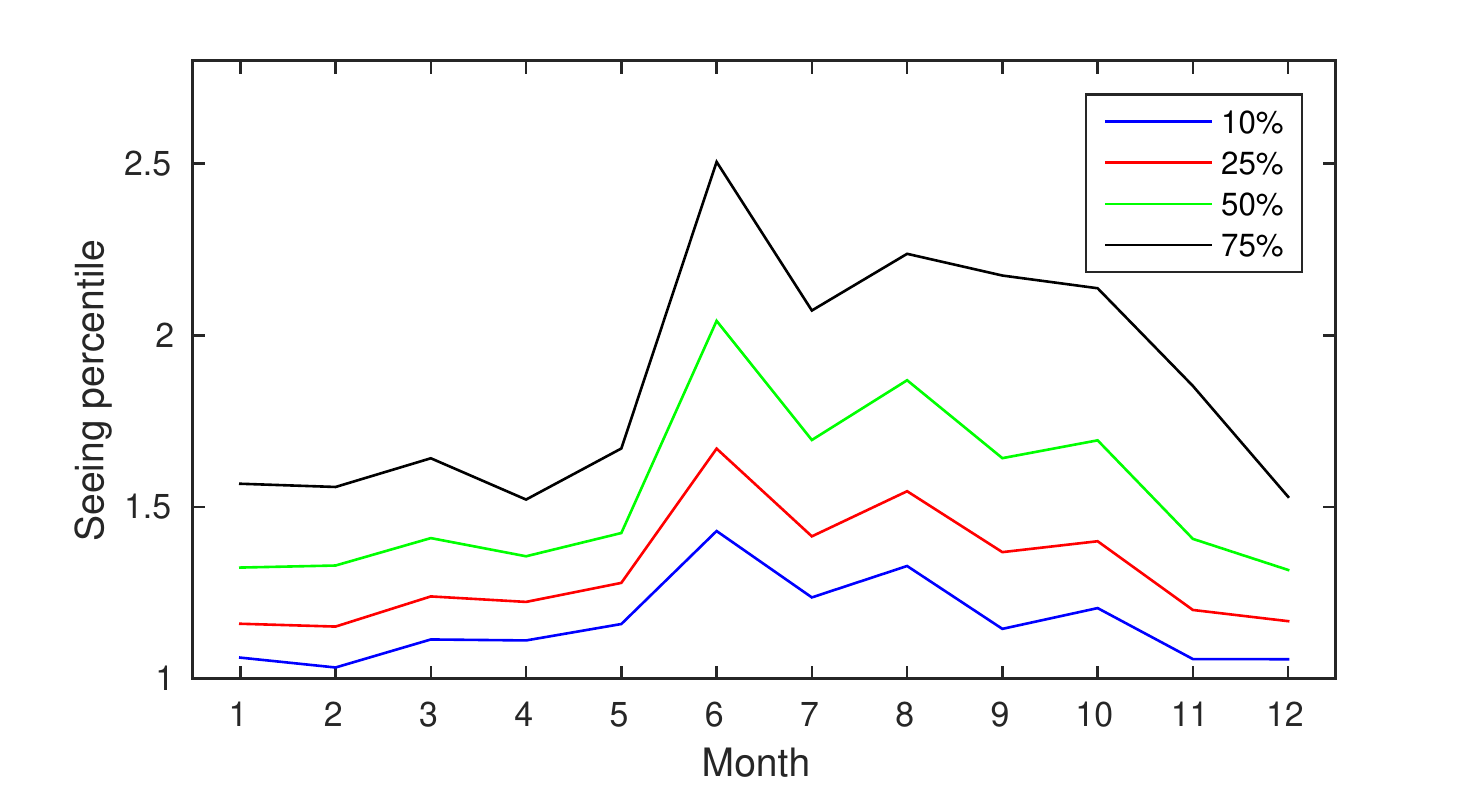}
\end{center}
\caption{Statistics of the seeing FWHM observed by TIGRE. The 10\%, 25\%, 50\%, and 75\% percentiles of the seeing distribution in each months are displayed.}\label{fig:seeing4_month}
\end{figure}

As is the case with many other weather parameters, the seeing also depends strongly on the season. The median values range from 1.4$''$ in January to 2$''$ in June (Fig.~\ref{fig:seeing4_month}). The monthly median is below 1.5$''$ between November and May. Furthermore, the seeing is better than 1.5$''$ for more than 60\% of the time and even below 1.2$''$ for 30\% between December and May. On the other hand, in June, only 18\% of the time the seeing is below 1.5$''$. The season with bad seeing starts in June and continues until October when the median seeing is 1.7$''$ and only 37\% of the time it is below 1.5$''$.

However, these measurements have some limitations. First, the seeing also depends on the wavelength, being smaller for redder wavelengths. Since the guiding camera has no color filter, the measurements also change with the star's color. Second, there is a residual aberration in the optics of the telescope. This aberration is apparent when the seeing is below 1.5$''$, so the lower values of the seeing measured by TIGRE ought to be considered as upper limits of the atmospheric seeing. Furthermore, the seeing of an observatory is usually measured with a differential image motion monitor (DIMM), which operates at a particular wavelength (commonly 5000 \AA) and is independent of tracking errors and instrumental effects. For these reasons, we have not compared our results with other observatories.

\begin{figure} 
\begin{center}
\includegraphics[width=12cm]{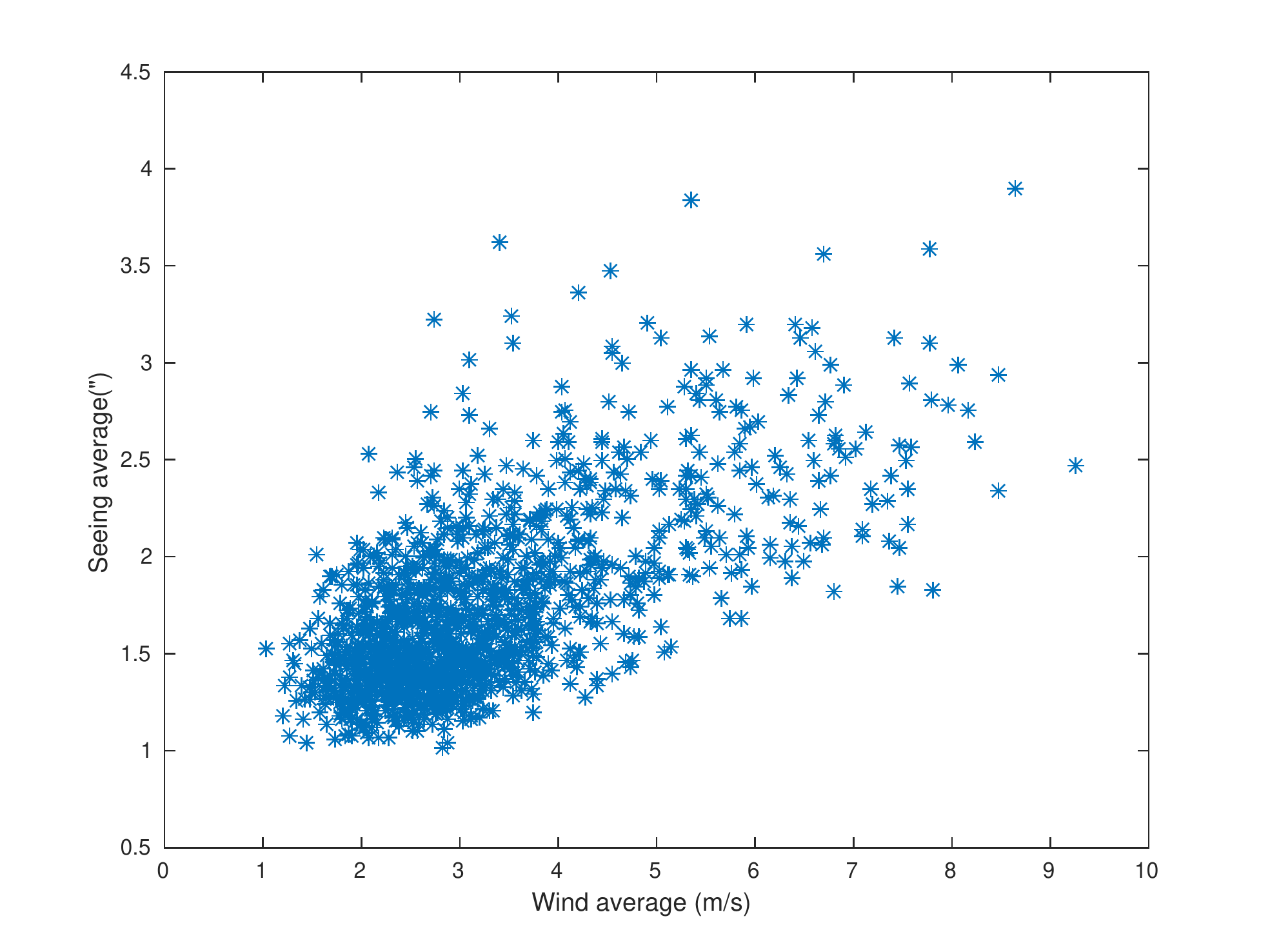}
\end{center}
\caption{Relation between the nightly averages of seeing and wind velocity. }\label{fig:seeing_wind}
\end{figure}

As mentioned above, we have noticed that the nights with strong wind always have poor seeing. Figure~\ref{fig:seeing_wind} shows the nightly-averaged seeing against the nightly-averaged wind velocity. Both parameters present a clear correlation, although this does not imply that the wind is directly causing the turbulence. In fact, usually at some point during windy nights, the wind velocity decreases to lower values, but the seeing remains high for several additional hours. It seems that it is the weather situation that produces strong winds also causes the atmospheric turbulence to be stronger and therefore causes more prominent seeing.

\subsection{Photometric nights}

Finally, we analyze here the photometric quality of the La Luz Observatory. Such an analysis is usually performed by estimating the photometry in a particular filter with a large aperture for many standard stars at different elevations. If the relation between the magnitude with the airmass is linear with small dispersion, the night can be considered photometric.

Since TIGRE is a spectroscopic telescope, it is very challenging to evaluate whether a night is photometric or not. The reasons are, first, the inherent difficulty of photometric calibration of echelle spectra, which is aggravated by the diameter of the fiber entrance: the fraction of photons feeding the spectrograph depends on the seeing as described above. Second, the guiding camera has no color filter, so the dependence of the instrumental magnitude on the star's color is strong and complex. This makes the guiding images in principle useless to estimate the photometric quality of the night. Third, we do not know the accurate values of the magnitudes of the stars; the great majority of them are not photometric standards, and many are actually variables.

As a consequence  another method must be used. Here, we calculate the statistic of the guiding camera photometry for each star observed in the night. The guiding cycle takes about 14\,sec, from which 10\,sec is the exposure time and the rest the time needed to readout and process the image and offset the telescope to the new position. Also, TIGRE spends most of the time in observations that last between 10 and 75\,min. Thus, there are between 40 and 300 flux estimates with the guiding camera for each star observed.

To evaluate the photometric quality of the night, we first calculate the standard deviation of the instrumental magnitude taken with the guiding camera for each star observed. For very long HEROS exposures, for which the airmass of the telescope has changed by more than 0.1, we corrected the magnitudes from extinction, linearly fitting the instrumental magnitude against the airmass and first removing the linear fit before the standard deviation is calculated.

We consider the night as photometric if the standard deviation of the instrumental magnitudes of the different stars is $<0.01$\,mag for at least 90\% of the time, and we have data for a minimum of 6 hours for each night. This method ensures that no clouds are obscuring the stars during the observations. However, this technique does not account for slow changes in the atmospheric extinction or variations depending on the direction of the observation.

During the years 2018 to 2021, we have obtained an average of a total of 40 photometric nights per year using this technique. The best months are December until February, with around 7 photometric nights per month. On the contrary, conditions are much worse in summer, when we have an average of less than two photometric nights per month between June and October. Since TIGRE's instrument is a spectrograph, the photometric quality of the observatory, despite not being excellent, does not change the capabilities of TIGRE significantly. Echelle spectra are complicated to be photometrically calibrated. The only possible consequence of a non-photometric night is thus to extend the exposures to reach the desired signal-to-noise ratio (see below).

\section{Performance and efficiency of TIGRE}
\label{Sect:perfor}

From a technical point of view, our goal is to expose as many good-quality spectra as possible. Here we describe how we have quantified the performance of TIGRE in these eight years and how we have pursued this goal. There are different aspects to be considered when talking about the performance:
        
\begin{itemize}
\item {\em Quality of the spectra:} The spectra obtained should meet the requirements of the astronomer. 
\item {\em Optical efficiency of the system:} From the incoming photons at the telescope's aperture, as many as possible should be detected in the spectrograph.
\item {\em Available time:} The astronomical (Sun below an altitude of -10$^\circ$) and meteorological conditions are constraints out of our control, and we have already discussed the weather above. 
\item {\em Technical issues:} Problems in the hardware are a major impediment for the observations, particularly in robotic telescopes, because usually no human intervention is directly possible when the issue appears. Some problems can be solved by the robotic operator and cause only a moderate amount of time loss. However, others are severe and may stop the activity of the facility possibly for months. 
\item {\em Operational efficiency:} This measures the fraction of the available time that is actually spent on exposures.
\end{itemize}

In the following subsections, we will describe in detail all of these subjects.

\subsection{Quality of the spectra}

We have a signal-to-noise calculator, yet it only gives reliable values when the atmospheric conditions are relatively good, and the actual conditions, under which the observations are carried out, 
are obviously unknown when the exposure is calculated.
To avoid taking too many underexposed spectra in case of poor conditions, it is of utmost importance to extend the
exposure time to reach the desired signal-to-noise ratio (SNR). Hence, to assure a well-exposed observation, the astronomer can request a minimum SNR at a particular wavelength and suggest the number of exposures in which the observation is divided and the exposure time of the individual exposures. Then, if necessary, TIGRE exposes longer to reach the required SNR, and so it adapts the exposure to the changing atmospheric conditions of seeing and transparency.

To estimate the SNR, the number of analog-to-digital units (ADUs) in a particular order in each channel ($CT$) is calculated once an exposure is taken\footnote{The wavelength ranges of the measured bands are $\sim5315\rm{\AA}-5355\rm{\AA}$ in the blue channel and $\sim7250\rm{\AA}-7300\rm{\AA}$ in the red channel}. TIGRE then estimates the SNR at the requested wavelength using these number of ADUs and the B-V color of the star.

The procedure that ensures that the requested SNR is reached, starts with the star acquisition. The instrumental magnitude of the star estimated by the guiding camera gives information about the transparency of the atmosphere. With this magnitude, the B-V color of the star, and the seeing, TIGRE can predict the exposure time necessary to reach the SNR. If this predicted time is longer than the one suggested by the astronomer, TIGRE calculates the exposure time for the individual exposures. If necessary, the robotic operator increments the number of exposures. After each exposure, $CT$ is calculated and compared with the expectation. After the predicted number of exposures are taken, the sum of $CT$ for all exposures is compared with the corresponding value of the requested SNR. Then, another exposure is taken if the target value for the $CT$ is not reached. The calibration of the relation between the SNR and both $CT$ and the instrumental magnitude of the guiding camera has been fitted using many stars of different spectral types. 

We re-emphasize that it is very easy for a robotic telescope to take data that do not reach the desired quality because no human evaluates the recently-acquired data and decides the necessary exposure time. With the method described here, we assure that TIGRE acquires good-quality spectra, regardless of the atmospheric conditions.

\subsection{Optical efficiency}

Without specific instruments, the determination of the system's absolute optical efficiency or only the reflectivity of the mirrors is impossible due to the temporal variation of the atmospheric conditions. Despite this impediment, we attempted to monitor the changes in the optical efficiency of TIGRE using alternative methods. For this, we considered the relative variations of the measurements of the guiding camera or the acquired spectra. The guiding camera photometry has the advantage that it does not depend on the seeing since one can calculate the magnitude with a large aperture. However, these data with large aperture were logged only over the last years. Also, we installed a new (and different) guiding camera in May 2019, and there is no easy method to compare the data taken with the two guiding CCDs. The use of the spectra, on the contrary, depends strongly on the seeing, but has the advantage that it is possible to compare the data taken now with data taken years ago. Also, the spectral measurements keep track of variations not only in the reflectivity of the mirrors of the telescope but also in the efficiency of the whole optical system, including the optical fiber, collimator, echelle grating, and cross-disperser of the spectrograph.

The main disadvantage of the spectral measurements, i.e. the dependence on the seeing, can be overcome by calculating how much light is lost outside the fiber entrance. We have calculated the flux losses using the difference of guiding magnitudes obtained with a large aperture (with a diameter of 6$''$) and an aperture with a diameter of 3$''$. This difference depends on the seeing and has been used to correct the data of the spectra.

Another problem of this method is that we do not know the precise magnitude of the stars. Many stars are variable, and for others, the magnitudes obtained from SIMBAD are not accurate enough. To solve this problem, we have used our sample of flux-calibration stars and their GAIA photometry \citep{2018A&A...616A...1G,2018MNRAS.479L.102C}\footnote{Our flux calibration stars are HR153, HR1544, HR3454, HR4468, HR7001, and HR8634}. This is a very small sample of stars, and one of them is observed every night whenever possible. Another advantage of this sample is that they all have similar colors, $-0.2<$B-V$<0.00$, which eases the analysis since the color dependence is negligible.

\begin{figure} 
\begin{center}
\includegraphics[width=18cm]{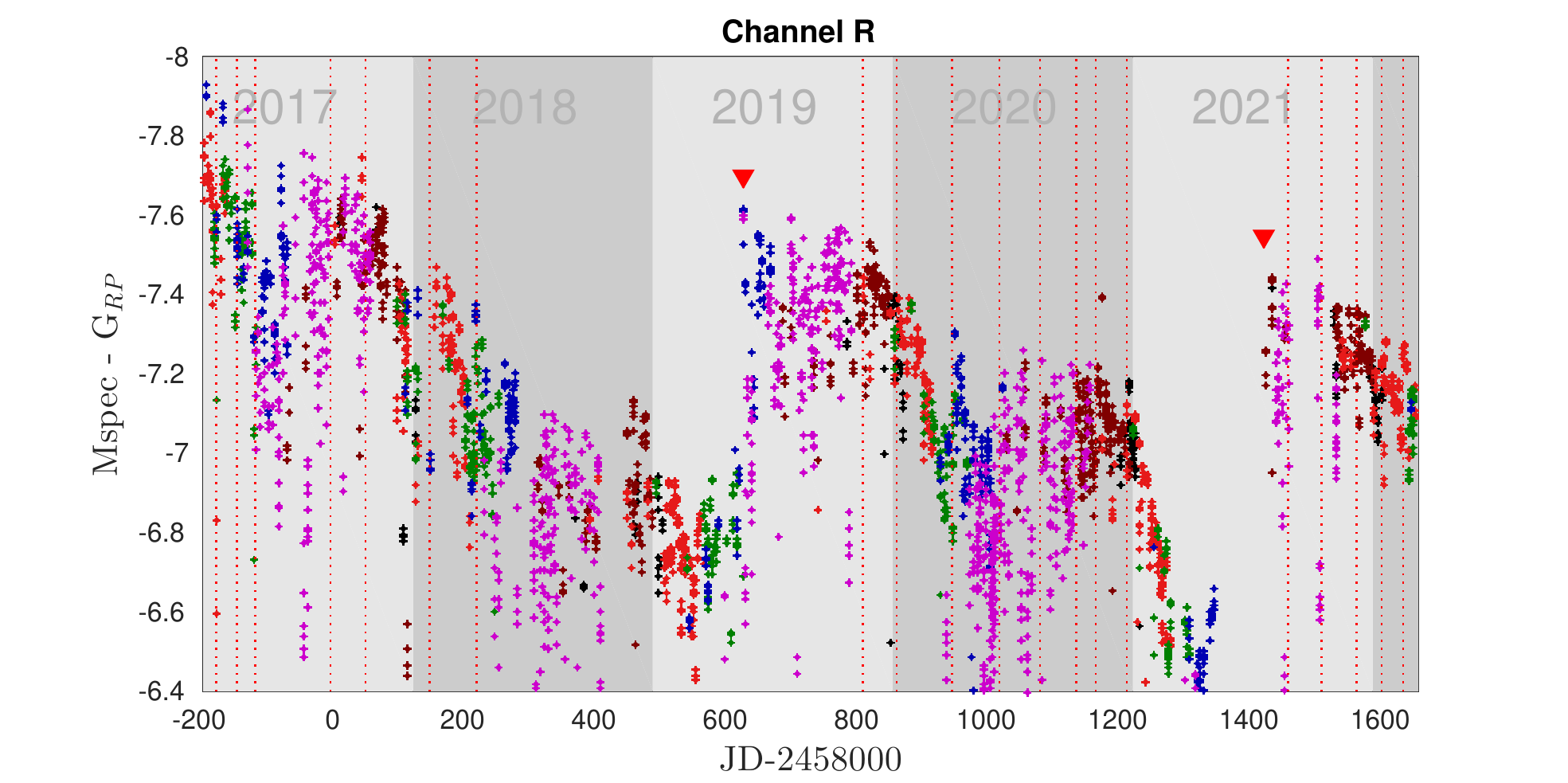}
\end{center}
\caption{Relative variation of the optical efficiency of the TIGRE-HEROS system. We show the difference of the GAIA magnitude and a seeing-corrected magnitude calculated with the HEROS spectral counts in a ca. 100 \AA-wide region around 7300\,\AA. The symbols with different colors represent the measurements of the six flux-calibration stars. The red triangles display the times when the mirrors were washed, and the red vertical dotted lines depict the times when the mirrors were CO$_2$-cleaned.}\label{fig:efficTelR}
\end{figure}

Thus, we calculated the magnitude $M_{spec}$, as a seeing corrected magnitude of the counts of the spectrum (as estimated above, $CT$) compared with the GAIA magnitude of all stars in the sample and presented their temporal variations in Fig.~\ref{fig:efficTelR}. There are significant variations from day to day due to the different atmospheric conditions. Still, when the conditions are excellent, the trends in the upper envelope are easy to notice. The big jumps in May 2019 and July 2021 resulted from the washing of the mirrors, and the red lines mark the time when a CO$_2$-snow cleaning was done.

In many telescopes, such as, WHT or VLT, the mirrors are cleaned with CO$_2$ snow \citep{1990SPIE.1236..952Z}. The cleaning process occurs when the snow quickly sublimates on the mirror surfaces and the particles of the mirror are blown away. This process is not abrasive and free of residues. We try to apply this cleaning once per month. 

We wash the mirror appoximately once per year with distilled water and a drop of dishwashing detergent. After washing the mirror, it is gently dried with absorbent paper.

The mirrors become dirtier with time. Also, the coating of the mirrors ages and deteriorates, so it is expected that the optical efficiency decreases with time, as is apparent in Fig.~\ref{fig:efficTelR}. Apart from this, washing is an excellent method to recover a large part of the reflectivity losses. Also, CO$_2$-snow cleaning reduces the losses in reflectivity. Figure.~\ref{fig:efficTelR} shows that the loss of optical efficiency is much faster when we could not clean the mirrors due to technical problems with the cleaning device. Curiously, the CO$_2$-snow cleaning is not always equally effective; in summer and autumn it is less effective than in winter and spring. However, if one looks carefully, the reflectivity losses are also higher in the winter and spring. This is because the dome stays open longer in the first months of a year, while this is not the case in summer. This, combined with the larger amount of dust in winter and spring due to less rain and dry soil, explains the more significant deposition of dust on the mirror of the telescope in these months; this new deposition of dust is what the CO$_2$-cleaning can remove.

\subsection{Technical issues}
\label{Sect:techIss}

TIGRE has suffered from several issues that limited or hindered its functionalities in these eight years. Here we can classify the problems regarding their severity:
               
\begin{itemize}
\item Minor issues that can be detected and solved by the error handler while the robotic operator continues working imply only a time loss of seconds or minutes. However, they may add to hours of time lost per month if they appear too frequently.
\item Other issues cause the robotic operations to stop completely, with time losses from hours to days. Examples of these are internet cuts and power failures. For some of these faults, human intervention is necessary to solve the problem.
\item Finally, severe issues can stop the observations for days to months. These are usually hardware failures that need the purchase of spare parts, and even the travel of a team from Hamburg to Mexico to repair the broken components. 
\end{itemize}

Although some issues are not avoidable, prevention helps to minimize the time lost. Some efforts have been made in this direction. First, we have continuously developed the error detection procedures and error handler. New issues appear occasionally, and our experience in the interactive solution is used to design a detection procedure and a solution that the error handler can perform. This is particularly important for the minor issues described above. Also, we have installed switchable sockets and remote-controlled USB hubs. With them, we can switch off/on any computer or device that has hung up. We also reboot the computers regularly, once per week, to avoid them from getting hung up when the different programs continuously run for a very long time.

Apart from the continuous improvement in the software, we have purchased several spare parts, in particular for the electronic cabinet. These include some relays, fuses, and other spare parts that the local staff can replace, and also a spare computer with its cards, as well as Profibus cards for the azimuth and elevation axes of the telescope. As a result, some of the severe technical faults we encountered were solved relatively quickly because we already had the spare parts at the telescope. This is particularly important for some essential components of the telescope electronics because they are no longer manufactured.

One of the long-standing problems for TIGRE is the power supply. Power outages do occur frequently, up to several times per month, and even more often in the rainy season. Since the power plant does not work very reliably, TIGRE closes and stops the robotic operation, and the system remains in idle status after an extended power failure (longer than five minutes), to avoid that the telescope stays open when the batteries of the UPS are out. In this last case, when the failure lasts several hours, the UPS does not restart automatically when the power returns, and the UPS needs human intervention to power up the system. However, the external power supply has improved in the last years: while we had $\sim$25 power cuts per year at night time between 2015 and 2019, the frequency reduced to $\sim$15 power cuts per year afterward. 

Figure~\ref{fig:lostHours} shows the number of night hours lost by technical issues in these eight years. To collect this information, we have extracted from the log files of TIGRE the amount of time when the robotic operator should be running, but it was not. In this statistic, the weather is not considered as long the system was running. The availability, as defined by \cite{syb2014}, is the fraction of the night time when the system runs without any issues. The availability of TIGRE has increased in the last years, thanks to the measures described above, except for the severe issues labeled in the figure listed below:

\begin{figure} 
\begin{center}
\includegraphics[width=12cm]{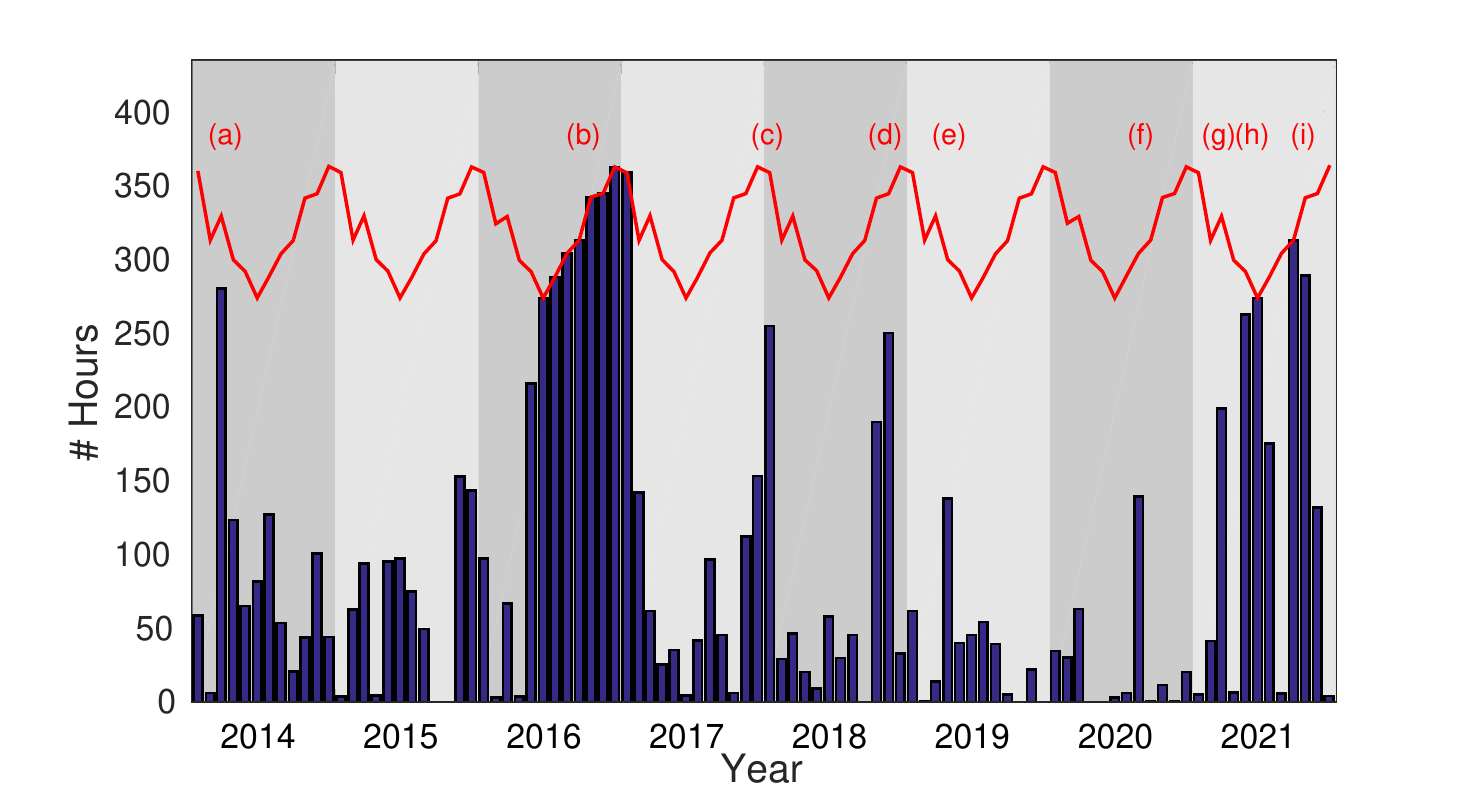}
\end{center}
\caption{Number of hour per month lost due to technical issues (bars). The red line display the total number of night hours. Specific technical problems caused the events labeled (a) to (i) as detailed in the text.}\label{fig:lostHours}
\end{figure}

\begin{itemize}                
\item[(a)] Hydraulic oil spillage.
\item[(b)] Mirror cell refurbishment and mirror aluminization.
\item[(c)] Failure in the calibration unit (flat field and ThAr).
\item[(d)] GPS antenna failure during a thunderstorm.
\item[(e)] Profibus card failure.
\item[(f)] GPS antenna failure during a thunderstorm.
\item[(g)] Failure in the hydraulic pump. 
\item[(h)] Fault in the focus-controlling card 
\item[(i)] UPS batteries in poor conditions.
\end{itemize}

\subsection{Operational efficiency}

We define operational efficiency as the fraction of the available time with good weather that is used in exposures. With this parameter we measure how well the available time of TIGRE is used. This parameter depends, of course, on the number of stars observed in the night. More stars with shorter average exposure times imply a lower operational efficiency because the overhead time, related to the number of stars observed, is larger. This overhead time includes the telescope's movement, the acquisition of the star, the beginning of the guiding, and the readout time. Technical issues as described above reduce also the operational efficiency.

This effect of the overhead time in the operational efficiency is unavoidable and comes from the requirements of the astronomer and the atmospheric conditions. However, other factors reduce the operational efficiency and are not desirable. Some of these factors are minor issues. After they are detected, the error handler often interrupts the robotic operation while solving them. Sometimes this involves restarting a device or even pointing the telescope and acquiring the star again.

Another factor is related to the weather conditions. Worsening of the weather often implies that the current observation is aborted. Unfortunately, the spectrograph CCDs do not allow to abort a running exposure and readout the collected electrons by then, which means that an aborted exposure is lost. To minimize this effect, we abort the exposures due to poor weather when the exposure is already performed by less than 70\% of the time. Besides, after a bad-weather period, when the conditions allow the observation, some time is needed to open the dome and initialize the telescope. which also reduces the efficiency.

To reduce the time lost due to the weather, we try to avoid observations with low chances to be performed successfully. This is accomplished by both the robotic operator and the scheduler.

\begin{itemize}
\item We avoid long observations in case of unstable atmospheric conditions. A straightforward approach is to select only targets whose observation takes not longer than the good-weather time. The latter is defined as the time since the last bad-weather event.

\item  The second possibility is to avoid the observations that require a minimum SNR and the calculated exposure time extends too much. Thus, observations that do not need much time are preferred, and hence they have a lower probability of being aborted because of poor weather or the end of the available time. The time necessary to reach the required minimum SNR is calculated to fulfill this condition. If the predicted time is much longer than the exposure time suggested by the astronomer, the observation is not performed. Therefore, the predicted exposure time should be:

\begin{equation}
\label{eq:lengthExpTime}
T_{exp,pred} < f * T_{exp,sugg},
\end{equation}

\noindent where $T_{exp,pred}$ and $T_{exp,sugg}$ are the predicted exposure time using the current conditions and the suggested exposure time. The factor $f$ depends on $T_{exp,sugg}$; it varies from $f=8$ for $T_{exp,sugg}<5$ min to $f=1.3$ for $T_{exp,sugg}>1.5$ hours. The predicted exposure time is calculated twice. First, the scheduler, knowing the last average conditions, i.e., extinction and seeing during the last observations, can predict the necessary total exposure time for the observation of each target to be successfully completed. In the next step, the robotic operator, using the information of the instrumental brightness obtained with the guiding star and the seeing, also predicts the necessary exposure time. When the predicted exposure time does not comply with Eq.~\ref{eq:lengthExpTime}, the scheduler does not select the target, or the robotic operator aborts the running observation and asks the scheduler for a new target.
\end{itemize}

In the future we plan to use weather forecasts to improve and optimize the selection of the targets to be observed, favoring some targets against others, when we have information about how the weather will be in the following hours or even days. 

\begin{figure} 
\begin{center}
\includegraphics[width=12cm]{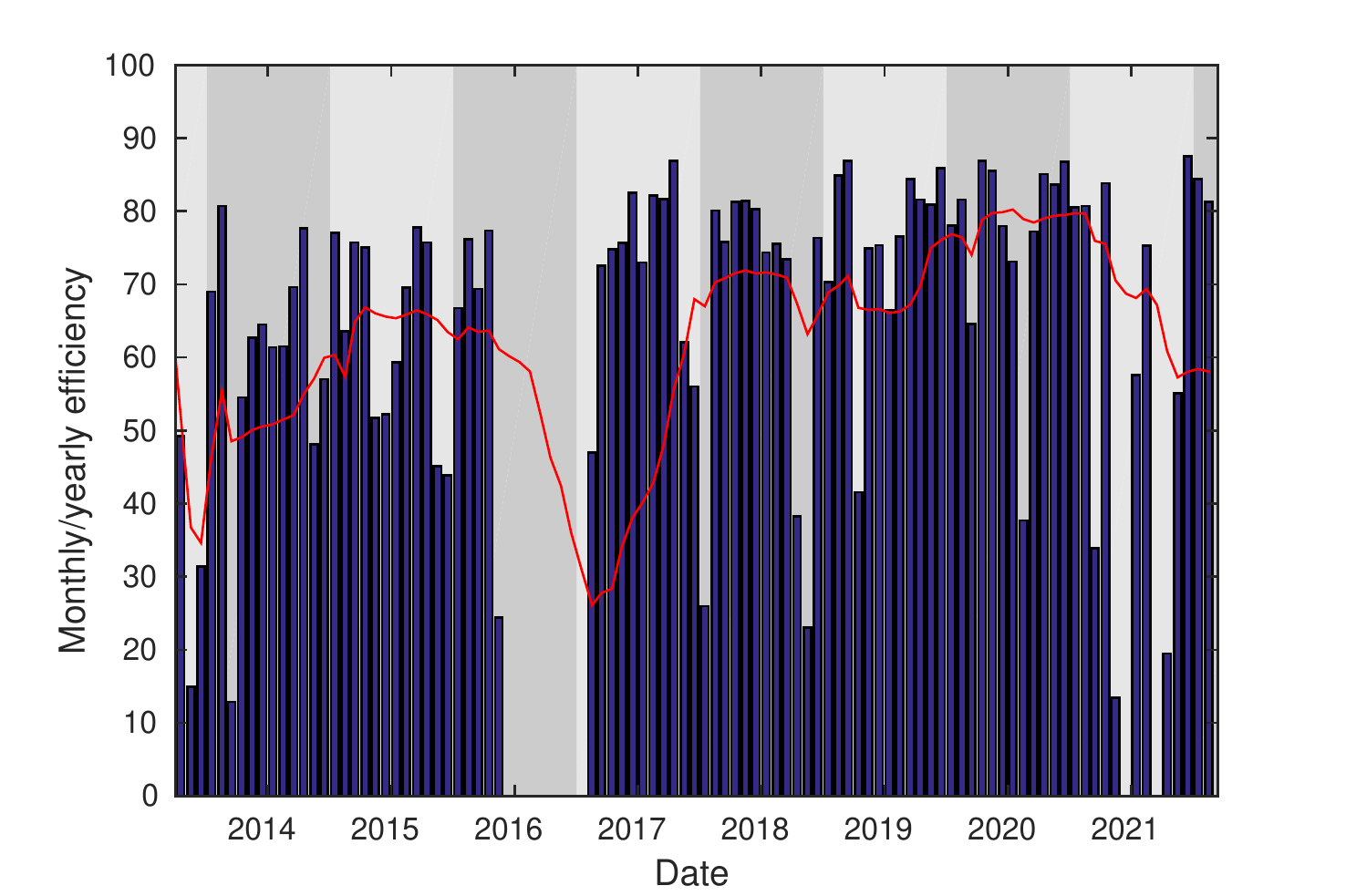}
\end{center}
\caption{Operational efficiency of TIGRE (bars). This is defined as the ratio between the total exposure time and the good-weather time. The red line shows the running 12-months average of the operational efficiency.}\label{fig:operEffic}
\end{figure}

Figure~\ref{fig:operEffic} shows the evolution of the operational efficiency in these eight years. Since 2017, the operational efficiency is high, usually higher than 70\% on average and often $>$80\%, except for the periods that coincide with technical problems as described in Sect.~\ref{Sect:techIss}. This performance has increased in the last four years due to the continuous improvement in software and hardware. In addition, the improvement in the operation described in this section has reduced the number of aborted spectra. Also, since more devices can be accessed and rebooted remotely when necessary, we can solve minor issues without waiting for the staff to drive to the observatory, reducing time losses. 

\section{Selected scientific results}
\label{Sect:Sci}

In this section we provide a short overview over (some of) the scientific output produced by TIGRE since its operational start in 2013.  In addition to quantitative measurements
we describe some of the scientific results that especially highlight the robotic capabilities of TIGRE.

\subsection{Collected science spectra}

\begin{figure} 
\begin{center}
\includegraphics[width=18cm]{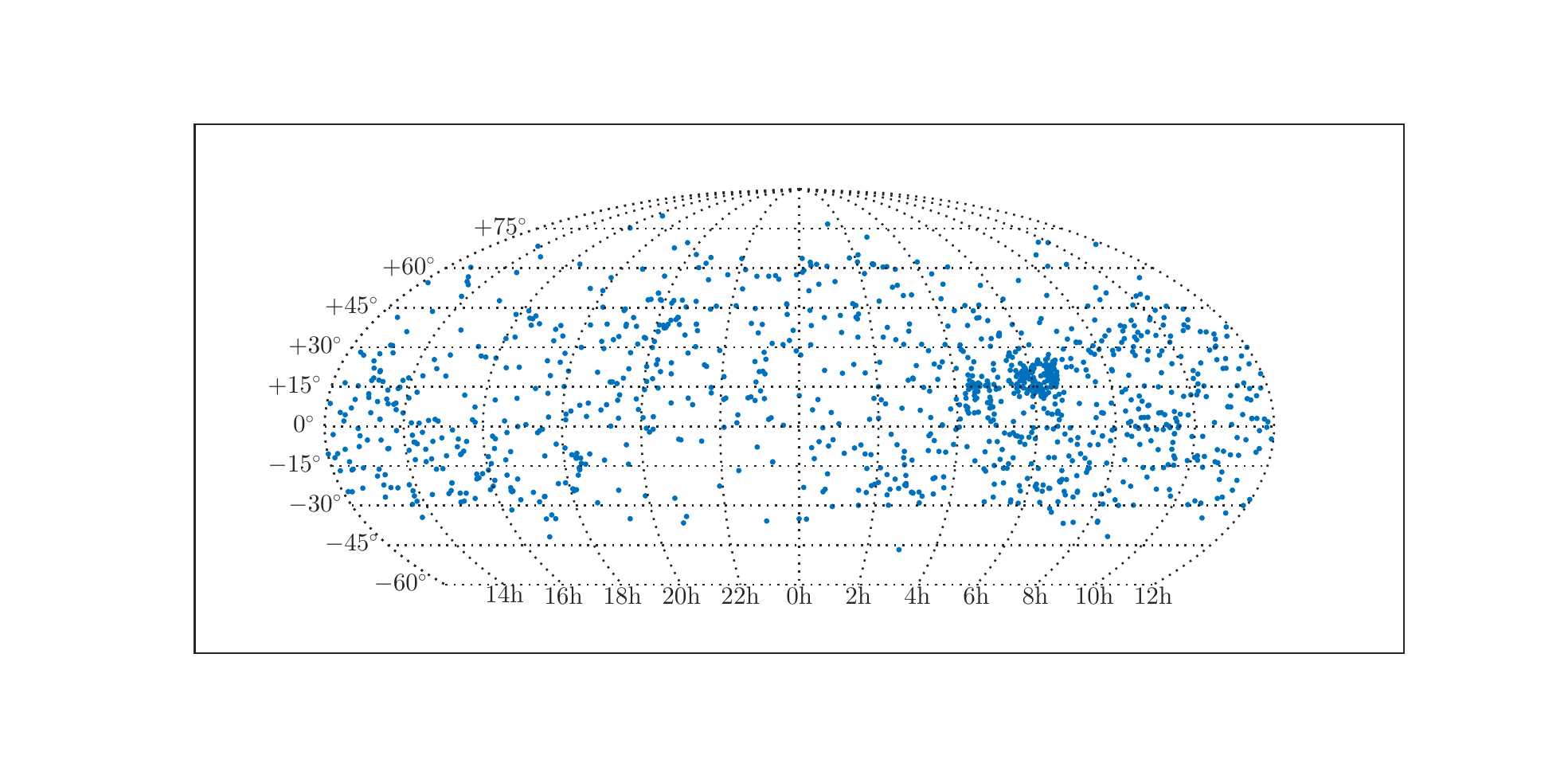}
\end{center}
\caption{Location on the sky in a RA-Dec projection of all stars observed by TIGRE}\label{fig:allStars}
\end{figure}

In the time frame 2013~-~2021 TIGRE has collected more than 48000 spectra of 1151 different sources with a total exposure time of more than 11000 hours.  To give an idea of where TIGRE observes,
we provide in Fig.~\ref{fig:allStars}  the distribution on the sky of all stars observed by TIGRE. 

\begin{figure} 
\begin{center}
\includegraphics[width=12cm]{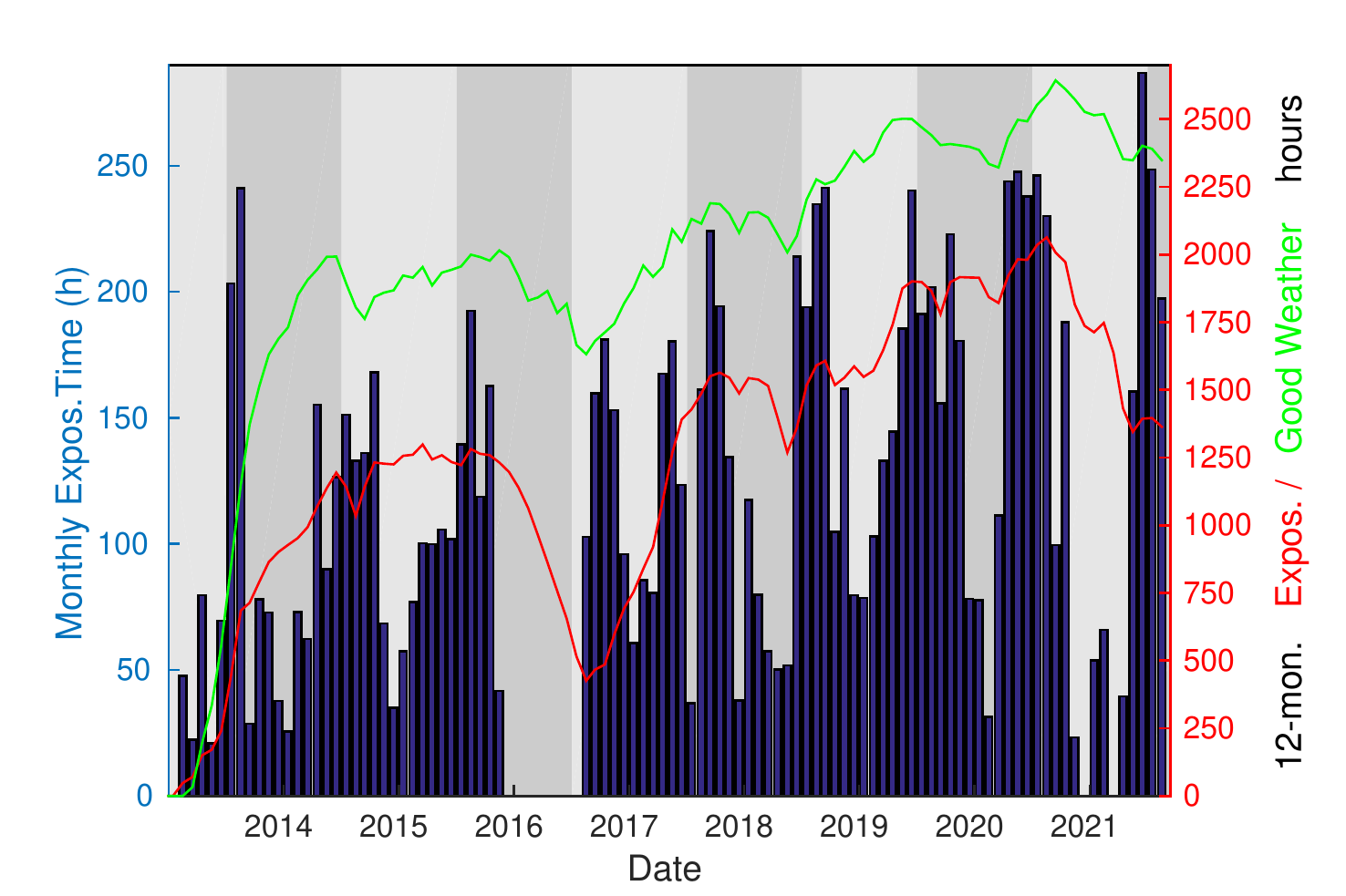}
\end{center}
\caption{The bars (left axis) show the monthly exposure time (in hours) since the start of TIGRE's operations. The lines (right axis) depict the 12-months total exposure time (red line) and the 12-months total number of good-weather hours (green line).}\label{fig:expTime}
\end{figure}

In Fig.~\ref{fig:expTime} we show the monthly total exposure time in scientific targets, and in Tab.~\ref{tab:totalSp} we provide  an overview of the number of nights 
with at least one scientific spectrum, the total number of spectra and total exposure time obtained with TIGRE.  As is clear from  Fig.~\ref{fig:expTime} ,
the total exposure time has continuously increased, except for 2021, when we were plagued with severe hardware failures. 
In 2019 and 2020, TIGRE obtained more than 1900\,h of exposure per year, which is a considerable improvement compared to the first years, 
when we could expose only $\sim$1200\,h. Also, the number of spectra taken has increased, from $\sim$5500 to $\sim$8000. However, this number is obviously less 
meaningful for the telescope's performance, because it also depends on the requirements of the astronomers and the stars' brightness. 
Fig.~\ref{fig:expTime} also shows the evident seasonal variations as well as the rising trend. In the summer, the total exposure time per month is reduced by a factor of
three due to the above described bad weather conditions between June and September.

\begin{table}
\caption{Number of nights with scientific spectra, total number of spectra and total exposure time for each year; note that in 2016 the mirror was re-aluminized, which explains the
lower numbers.}
\centering
\begin{tabular}{l r r r}
\hline\hline
Year & Nights & Spectra & exp. time(h) \\
\hline
2013  &   60   &  1692     &    240  \\
2014  &  211   &  5176     &   1156  \\
2015  &  242   &  5870     &   1233  \\
2016  &  109   &  2297     &   655   \\
2017  &  239   &  6428     &   1391  \\
2018  &  238   &  5778     &   1359  \\
2019  &  295   &  8738     &   1901  \\
2020  &  308   &  7455     &   1990  \\
2021  &  187   &  5064     &   1393  \\
\hline
Total &  1889  &  48498    &   11318 \\
\hline
\end{tabular}
\label{tab:totalSp}
\end{table}

\begin{figure} 
\begin{center}
\includegraphics[width=12cm]{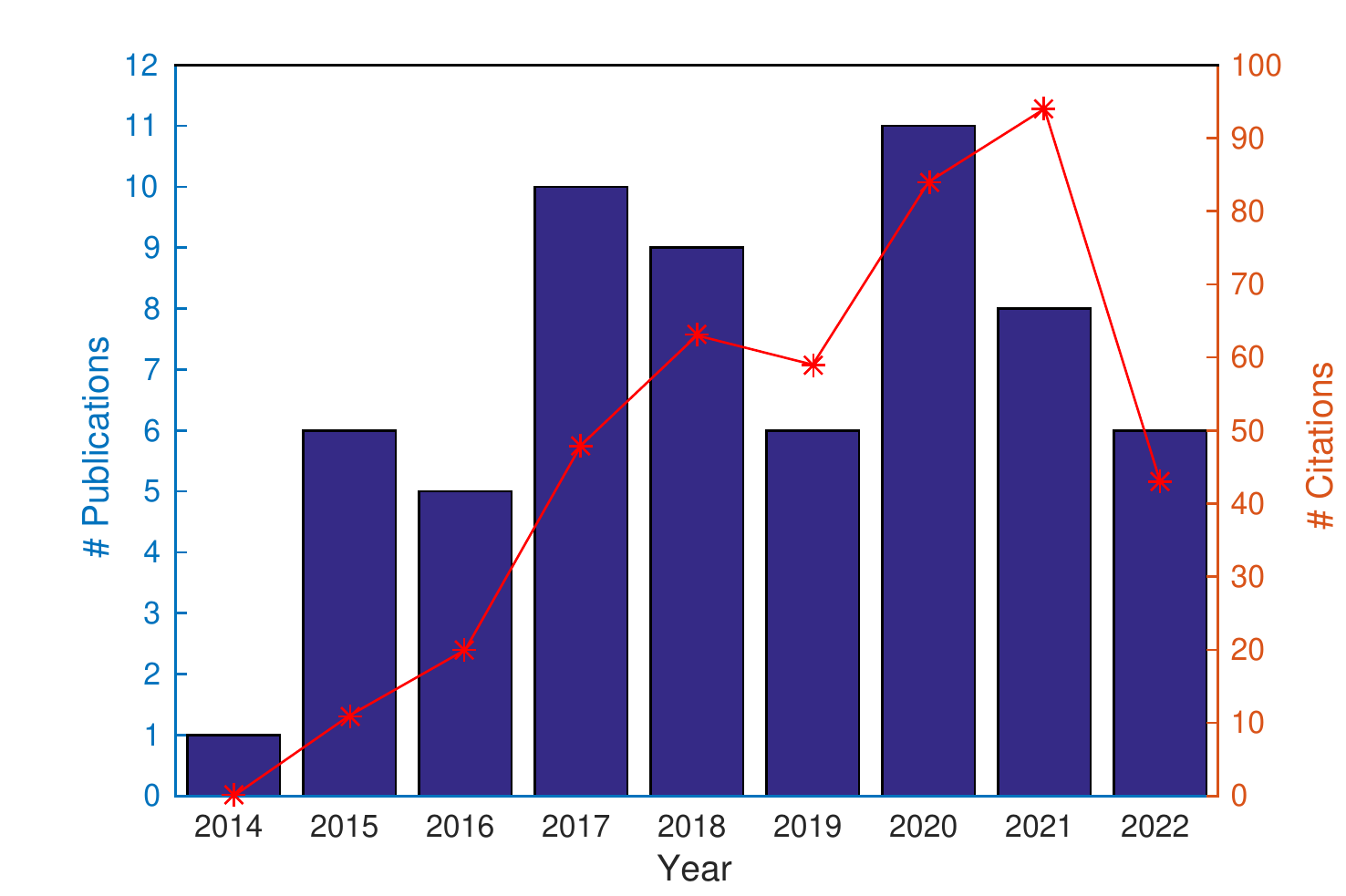}
\end{center}
\caption{Yearly number of TIGRE related publications (blue bars) and citations (red asteriks).}\label{fig:pubs}
\end{figure}

\subsection{Publications}

Publications in recognized journals are clearly an indicator of the success of any telescope facility.   We are maintaining
a web site under \url{https://hsweb.hs.uni-hamburg.de/projects/TIGRE/EN/hrt_publication.html}, which is updated
once or twice a year and lists all publications involving TIGRE data.  At the time of writing (March 2022) we
have 62 such publications.  In Fig.~\ref{fig:pubs} we provide a graphical display of the temporal evolution of the
TIGRE publications and citations.

\begin{figure} 
\begin{center}
\includegraphics[width=17cm]{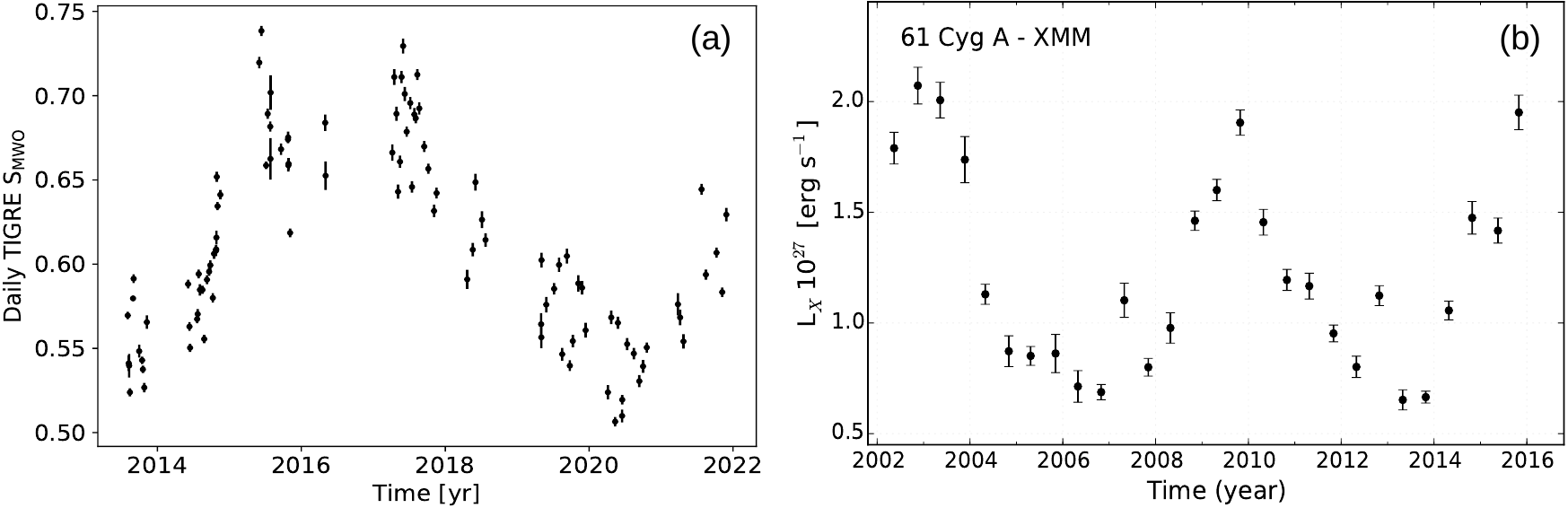}
\end{center}
\caption{(a, left panel): TIGRE S-index time series of the active star HD~201091; note the clear cyclic
variability in the S-index. (b, right panel): Evolution of X-ray luminosity derived from {\it XMM-Newton} X-ray monitoring of HD~201091 between 2002 and 2016; note the cyclic variability of the recorded X-ray emission (courtesy J. Robrade).}\label{fig:HD201091_sind_xray}
\end{figure}



\subsection{Stellar parameters and activity of cool stars}

Long term monitoring is the basic requirement for the observational study of stellar activity cycles analogous to the
well-known eleven year cycle of the Sun.  However, stars remain -- almost always -- spatially unresolved,
and furthermore, lower activity stars like the Sun are typically faculae-dominated rather
than spot-dominated.  As a consequence, cyclic variations are easier to study in chromospheric
emission lines using moderate- to high-resolution spectroscopy rather than broad band photometry.
The lines of choice are the deep Ca~II H\&K absorption lines, already recognized by
\cite{fraun1817} during his early solar studies.   The cores of the absorption lines at 3968.469\,\AA\, 
(Fraunhofer's H line) and 3933.663\,\AA\, (Fraunhofer's K line) exhibit very low residual
photospheric flux, thus allowing the detection of the overall much fainter chromospheric emission
against a rather low photospheric background.  

\cite{eber1913} were the first to recognize reversals of the Ca~II H\&K line cores in the spectra
of a few stars observed by them and  realized the potential offered by these line reversals
to study stellar cycles, writing: "It remains to be shown whether the emission lines of the star
show a possible variation in intensity analogous to the sun-spot period".  However,
a systematic study of these line core variations in a larger sample of stars over an
extended period of time was taken up only much later, in the 1960ies,  by O.~Wilson in his
Mount-Wilson H\&K program. Rather than using photographic plates as done by  \cite{eber1913},
O.~Wilson employed a specially designed H\&K photometer, which used far more sensitive
photoelectric photometers to measure the flux in two triangular spectral bands with a width
of a little over 1~\AA\, centered on the wavelengths of the H and K lines.   To remove
effects of seeing and variable atmospheric emission, the measured line core fluxes are
provided relative to the flux in two nearby continua redwards and bluewards of the
H\&K lines as so-called S-indices. The first results
of this monitoring program were reported by \cite{wilson1978}, where he
(tentatively) concludes that:" (1) no stellar chromospheres are likely to be constant in time; (2) short-term fluctuations tend to 
increase in size with average flux; (3) cyclical variations occur with periods ranging from about 7 years to probably at least 
twice as long; (4) the stellar cycles observed in H and K flux should be regarded as evidence for analogs of the solar cycle; 
and (5) the incidence of complete or probable partial cycles increases toward later spectral types."

The more or less final results (although some observations were carried out later) of the Mount-Wilson program
were presented by \cite{baliunas1995}, however, it was becoming clear at the time that long-term monitoring of
late-type stars could be (and should be) carried out in a far more efficient and economic way by using robotic 
telescopes (thus saving on man power) and using high-resolution spectrographs, which made any
individual settings (to adjust e.g. for the individual star's radial velocity) unnecessary.  Furthermore,
an automated  data-reduction pipeline reduces the spectra efficiently and reliably and produces,
for example, S-indices or other activity indices quickly and automatically.   Furthermore, spectrographs
make full use of the whole spectral range covered, while the H\&K photometer ``threw away'' a large
part of the actually recorded spectral information.

TIGRE encompasses all these features, but it can -- of course -- also provide S-indices which
are calibrated against the extensive Mount-Wilson S-index data base.  As an example, 
in Fig.~\ref{fig:HD201091_sind_xray} we show the S-index obtained from the TIGRE spectral time series of the
active star HD201091 (= Cyg~61 A, spectral type K5V), which very beautifully demonstrates the
cyclic nature of the chromospheric emission from HD~201091 with a cycle period of 
about seven years and nicely confirms the Mount-Wilson results
on the same star reported by \cite{wilson1978}.  HD~201091 is remarkable since we monitored the X-ray 
emission of this star using {\it XMM-Newton}. Due to operational constraints, X-ray monitoring is far less
dense (typically only two observations per year) than optical monitoring, however, the long-term X-ray light curve of HD~201091, displayed
in Fig.~\ref{fig:HD201091_sind_xray} shows a clear cyclic variability with the same period as the
chromospheric emission.   Once TIGRE went online in 2013, we accompanied the {\it XMM-Newton}
observations with simultaneous TIGRE observations, and indeed, the chromospheric emission
strongly increases between 2014 to 2016 in line with the increasing X-ray emission.

Furthermore, the comparison of observed spectra with synthetic spectra from model atmospheres yields the main physical parameters of the observed star, namely effective temperature, gravity, metallicity, rotation velocity and turbulent broadening parameters. TIGRE and the very homogeneous quality of its HEROS spectra, monitored each night as part of the reduction pipeline and regular calibration procedures, has proven very useful for such work on larger samples. The consistency of such an analysis for a whole sample depends a lot on the reproducibility of the spectral resolution, because a false consideration of the latter leads to confusion over the velocity parameters and gravity broadening, indirectly even affecting the effective temperature and metallicity determination, see \cite{2021MNRAS.501.5042S}. There is a good potential for characterizing larger samples of cool stars consistently, using TIGRE spectra, as would be of interest for, e.g. exoplanet host-stars and other samples of particular interest.

The reliance on homogeneous resolution in a sample of spectra is an especially sensitive issue with cool stars, and giants in particular, where spectral lines are intrinsically sharp  (see Rosas Portilla et al. 2022, MNRAS, in print). A sample of TIGRE spectra for over 30 cool giants have already helped the revision of the physical relations behind the Wilson-Bappu effect, for which we measured the Ca II K emission line width in the blue channel spectra and used the red channel, which is providing a less line-crowded and so easier to analyze spectral range, for the physical parameter analysis.

\subsection{Hot stars}

The instrumental stability and scheduling flexibility of TIGRE are major assets that are beneficial to a number of different studies. To illustrate this point, let us consider the case of 
a peculiar class of massive stars, the $\gamma$~Cas stars. These objects are Be stars, i.e.\ rapidly rotating B-type stars surrounded by a Keplerian circumstellar disk 
containing material ejected by the star. Whilst the majority of the Be stars display a comparatively weak ($\log{L_{\rm X}/L_{\rm bol}} \sim -7$) and soft ($kT \sim$ 0.2 -- 0.6\,keV) 
X-ray emission, the so-called $\gamma$~Cas stars feature a bright ($\log{L_{\rm X}/L_{\rm bol}} \sim -6$ to $\sim -4$) and unusually hard (kT $\sim 12$\,keV for $\gamma$~Cas) 
X-ray emission \citep{Smi16}. To date, about two dozen objects are known to belong to this category \citep{Naz20}. Various scenarios have been proposed to explain their peculiar 
X-ray properties, including accretion of the Be disk material by a compact companion, wind interactions with a hot stripped helium star companion, or magnetic interactions between 
the Be star and its disk. In this context, investigating the multiplicity of $\gamma$~Cas stars is crucial to discriminate between the various hypotheses. Over the last eight years, 
we have used TIGRE to monitor the optical spectra of a sample of $\gamma$~Cas stars. Thanks to its robotic mode, TIGRE enabled us to collect well-sampled time series of 
spectra which were used to perform a systematic radial velocity monitoring of these stars \citep{Naz22}. This study showed that the properties of the known and newly found 
$\gamma$~Cas binaries do not significantly differ from those of other Be binary systems and that companions are of low mass (0.6 -- 1\,M$_{\odot}$). 

Beside the 
multiplicity investigation, the TIGRE spectroscopic monitoring allows to characterize the changing properties of the Be disk via the strength and shape variations of a variety of 
optical (H\,{\sc i} Balmer series, He\,{\sc i}, Fe\,{\sc ii}) and near-IR (H\,{\sc i} Paschen series) emission lines arising within the disk. Indeed, establishing the link between the Be 
disk properties and the peculiar X-ray emission is an important key towards understanding the origin of the hard X-ray emission. A coordinated monitoring with TIGRE and 
the {\it Swift} X-ray telescope over three orbital cycles of $\pi$~Aqr unveiled no correlation between the X-ray emission and either the orbital phase or the H$\alpha$ emission 
strength \citep{Naz19}. 

\begin{figure}
\begin{center}
\resizebox{9cm}{!}{\includegraphics{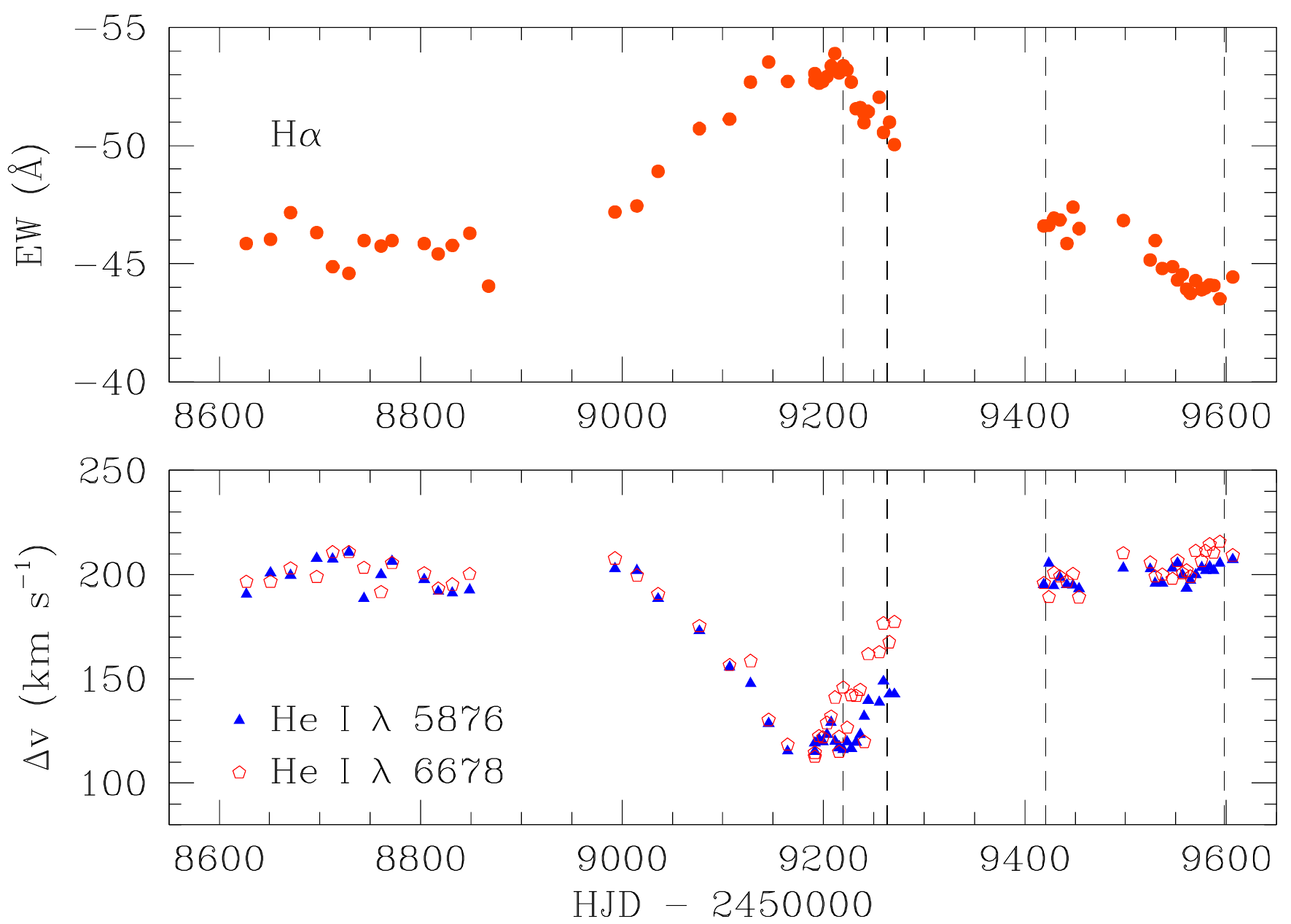}}
\caption{TIGRE monitoring of $\gamma$~Cassiopeiae between May 2019 and January 2022. The top panel illustrates the equivalent width (EW) of the H$\alpha$ emission line as a function of time, whilst the lower panel displays the velocity separation ($\Delta v$) between the violet and red peaks of the double-peaked emission components of the He\,{\sc i} $\lambda$~5876 and $\lambda$~6678 lines. The outburst around HJD\,2459180 (January 2021) is clearly seen in EW(H$\alpha$) and via a significant change in $\Delta v$, indicating that the formation region of the He\,{\sc i} lines had expanded significantly. The dashed vertical lines indicate the times of our {\it XMM-Newton} ToO observations.\label{gCas}}
\end{center}
\end{figure}

Moreover, the TIGRE spectroscopic time series allow us to trigger target of opportunity observations with the {\it XMM-Newton}, {\it Swift} or {\it Chandra} 
X-ray observatories when the Be disks undergo a major transition, i.e.\ either an outburst or dissipation. Once these X-ray observations are scheduled, TIGRE further offers the 
possibility to collect optical spectra nearly simultaneously with the X-ray data. This strategy has been applied with success to the cases of a disk fading 
event in HD~45314 \citep{Rau18}, a cyclic outburst of the normal (i.e.\ not $\gamma$~Cas) Oe-star HD~60848 \citep{Rau21}, and most recently an outburst of  $\gamma$~Cas
 itself (see Fig.\,\ref{gCas}). The multi-wavelength data collected in those campaigns favour scenarios where the X-ray emission arises from 
 the innermost parts of the Be disk, unlike what one would expect if the X-ray emission were due to interactions with a companion. 

\subsection{Transient objects: Novae and Supernovae}

A further strength of TIGRE is its quick response time for observations of targets of opportunity.  Once
we receive information about a newly discovered event such as a nova or supernova, 
we can usually observe the respective target already during the next night weather and sky position 
permitting.  As a consequence we have been able to carry out
successful observation campaigns of several supernova and nova targets.

Needless to say, given TIGRE's telescope aperture of 1.2~m, we can observe only the brighter novae and
supernovae, yet during the past eight years of TIGRE observations, we were able to observe two 
bright supernova events.
The very bright supernova SN~2014J was observed with a long time series during the
dry winter season in January and February of 2014 \citep{jack15b}.  With TIGRE
we took in total 43 spectra during a period of over two months, and 
this densely sampled time series helped identify the features that cause the secondary maximum
in the near-infrared light curves of supernovae of type Ia \citep{jack15}.
We also took six spectra of the Supernova SN~2017eaw. However, this supernova was too
faint for TIGRE observations, and we only obtained spectra with an acceptable S/N in the red channel. 
Thus, supernovae are in general difficult targets for a small sized telescope like TIGRE, but 
given bright enough SNe (such as  SN~2014J) the spectral time series that can be obtained are 
quite unique.

Observations of galactic classical novae are much easier for TIGRE, since these events are 
much brighter and also more numerous.  Our first TIGRE nova (i.e., the Nova V339 Del) 
was already observed in 2013 during the first months  of TIGRE observations \citep{novadel}.
The longest nova time series was taken for Nova V5668~Sgr, for which we were able to obtain more than  200 spectra
during a period of about two years. 
This nova of the type DQ Her showed rapid variations in both the light curve and spectra during the first 100 days
after the outbreak.   With TIGRE we were able to study the rapid changes in the spectral features of several lines
during that phase in detail \citep{novasgr}.
Another time series was collected of Nova V659~Sct \citep{novasct}, for which we obtained eight spectra.
A further nova that we observed was Nova V1112~Per, for which we obtained in total 34 spectra, and 
another twelve spectra were obtained of the Nova V6593~Sgr. 
The most recent observation of a TIGRE spectral time series are 16 spectra of the 2021 outbreak of RS~Oph.
Thus, we took in total over 300 spectra of six galactic novae during the eight years of TIGRE observations.
Another aspect of the intermediate resolution spectra of TIGRE is that it allows us also to study 
interstellar absorption lines and diffuse interstellar bands in the nova spectra \citep{novais}. 

\begin{figure}
\centering
\includegraphics[width=0.5\textwidth]{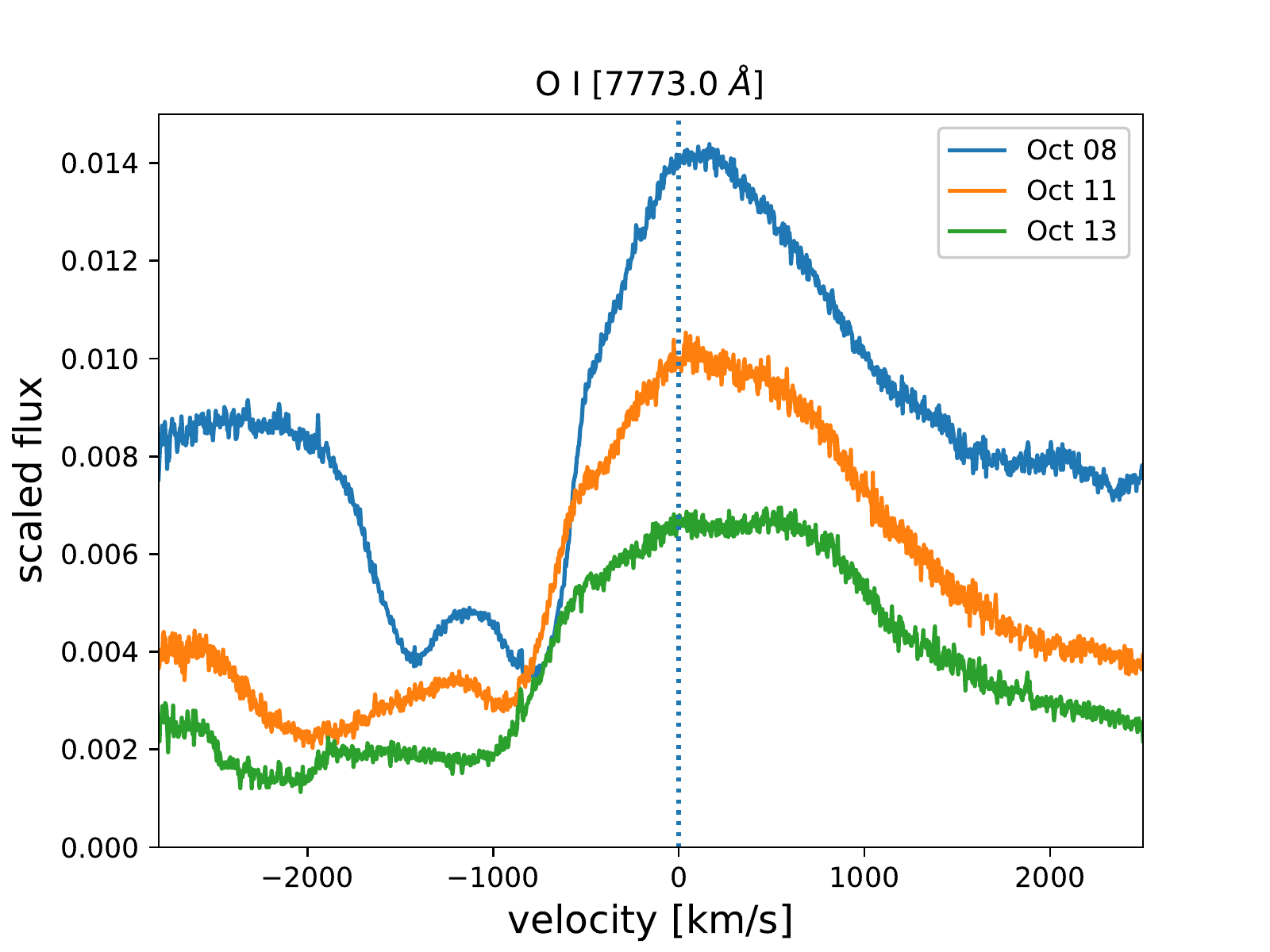}
\caption{Three spectra of Nova V6593 Sgr in the O\,{\sc i} line at 7773.0~\AA\, observed in 2020; note the rapid changes in the recorded line profiles,
indicating the increase in expansion velocity.}
\label{fig:novae}
\end{figure}

To demonstrate the advantages of the TIGRE spectral time series observations 
we present as an example in Fig~\ref{fig:novae} 
the evolution of the very common spectral line of O\,{\sc i} at 7773.0~\AA\ of Nova V6593~Sgr. 
There are two absorption features at about $-800$ and $-1500$~km~s$^{-1}$
in the spectrum from October 8th in 2020. 
These two features move during the following days to higher expansion velocities.
This characteristic evolution was also observed in other novae.
The robotic operation and the intermediate resolution of the HEROS
spectrograph make TIGRE an ideal telescope for the observations of time series 
of novae and supernovae, and we are eager to observe more supernovae and novae
targets of opportunity in the coming years.

\section{Summary and outlook}

In the present paper we have described our experiences collected in eight years of operating TIGRE in a fully robotic mode, and here we attempt to summarize our main findings.  We can state that our international TIGRE collaboration was really fruitful scientifically and extremely helpful operationally; having one of our partners on site, was essential for all logistic support and for trouble shooting in the cases of unexpected failures.  And finally, at the La Luz site there is manpower (``vigilantes'') who could be contacted in the case of absolute emergency.

Needless to say, a reliable internet connection is a ``conditio sine qua non'' for any robotic telescope, and it took some measures and time for TIGRE to arrive at that point.   Reliable power supply is a further essential requirement, and the power supply at the La Luz Observatory site has definitely room for improvement. During the summer time thunderstorms are a constant source of concern, and further, the nominal power frequency of 60~Hz is not that stable.   We have hardware to compensate for power failures in the form of batteries and a Diesel generator, but the reliability of the system can certainly be improved.  On the long run, the use of solar power could make TIGRE self-sufficient power-wise, at least to a large extent.

While we can counteract at least some power supply problems ourselves, little can be done as far as dust contamination is concerned.  The telescope is already mounted on a rather high pier (cf., Fig.~1) to avoid the worst of dust, however, the main reduction of dust production occurred as a consequence of the nearby road being paved, which was actually beyond our control.

TIGRE has only one scientific instrument, i.e., its HEROS spectrograph.  Although there are quite a few robotic telescopes in operation, only very few of them deal with spectroscopy;  this is clearly TIGRE's unique selling point. While the spectral resolution of HEROS is ``only'' a little over 20000, we feel that this resolution is entirely sufficient to address a wide range of scientific questions.  Obviously, for some applications, such as high-precision RV work or Doppler imaging, HEROS, is not well suited, however, one also has to keep in mind that the phase space for such observations is quite small for telescopes in the 1m class, and, for example, meaningful Doppler images can be produced only for a relatively small number of sources. What has proven beneficial, is the dual arm nature of HEROS.  It provides, first, redundancy when one channel fails, and, second, allows the independent operation of the two channels which is very advantageous, for example, for rather red objects with small emissions in the Ca~II H\&K line cores with different exposure times in the red and blue channels.

Scientifically one may ask the question of what kind of role a (robotic) telescope of the 1m-class can play in the epoch of 10m (and in the future even larger) telescopes. In our view, our hitherto TIGRE record shows that indeed a scientific niche can be found for a robotic facility that is capable of producing valuable and visible scientific results. While essentially all of the TIGRE observations could in principle also be carried out by larger telescopes, the operation of larger facilities is far more costly, so in practice the spectral time series produced by TIGRE are unique.  Similar considerations apply to the observations simultaneously carried out by TIGRE with space-based observatories such as XMM-Newton or eROSITA; TIGRE can carry out such observations in a rather straight-forward and uncomplicated fashion in an automated manner.  For all practical purposes TIGRE works like a space-based robot, except that the conditions that TIGRE encounters are -- at least sometimes -- more adverse than that of a facility based in space. Thus, with TIGRE, the participating university institutes, which are all quite small compared to research institutes positioned outside the university system, have the opportunity to obtain access to very unique data sets.

For the TIGRE system the practical magnitude limit is between 8 - 10~mag depending on the specific requirements of the observer, fainter objects (such as the SN 2014J) can of course also be observed, however, the investment of observing time becomes excessive large. In this context we mention the ongoing TESS (photometric monitoring for exoplanet search) and eROSITA (X-ray all-sky survey) missions as well as the upcoming PLATO mission (photometric monitoring for exoplanet search) which are and will be yielding thousands of ``interesting'' brighter sources which require optical follow-up.  As a matter of fact, the spectroscopic follow-up is already now the bottleneck for TESS, and the same will apply to PLATO.  Thus it appears that also in the coming years there will be a strong scientific demand for spectroscopic robotic telescopes  like TIGRE.

\section*{Conflict of Interest Statement}

The authors declare that the research was conducted in the absence of any commercial or financial relationships that could be construed as a potential conflict of interest.

\section*{Author Contributions}

All authors listed have made a substantial contribution to the work and approved it for publication.

\section*{Acknowledgments}

We acknowledge the  continued support by various partners who helped to realize and operate TIGRE. In the first place, this is the University of Hamburg, which gave support in terms of funding, manpower and workshop resources. Furthermore, grants from  the  Deutsche  Forschungsgemeinschaft  (DFG)  in  various funding lines are gratefully acknowledged, as well as travel money from both the DFG, DAAD and Conacyt in several bilateral grants. The Li\`ege contribution to the TIGRE is funded through an opportunity grant from the University of Li\`ege. The Universities of Guanajuato and Li\`ege and the Mexican state of Guanajuato shared the funding of the infrastructure required by TIGRE at the La Luz site, and there is continued support by the University of Guanajuato in terms of manpower and running costs of the facilities. YN and GR acknowledge support from the FNRS-FRS (Belgium). We thank Alexander Hempelmann for his invaluable dedication and contribution to this project.

\bibliographystyle{Frontiers-Harvard} 

\bibliography{tigre_roboticTelescopes}





\end{document}